\documentclass[twocolumn]{aa}
\usepackage{graphicx}
\usepackage{txfonts}
\begin{document}
\title{An optimal hydrodynamic model for the normal Type IIP supernova 1999em}

\author{Victor P. Utrobin\inst{1,2}}

\offprints{V. Utrobin, \email{utrobin@itep.ru}}

\institute{
   Max-Planck-Institut f\"ur Astrophysik,
   Karl-Schwarzschild-Str. 1, D-85741 Garching, Germany
\and
   Institute of Theoretical and Experimental Physics,
   B.~Cheremushkinskaya St. 25, 117218 Moscow, Russia}

\date{Received 21 July 2006 / accepted 22 August 2006}

\abstract{
There is still no consensus about progenitor masses of Type IIP supernovae.
}{
We study a normal Type IIP SN 1999em in detail and compare it to a peculiar
   Type IIP SN 1987A.
}{
We computed the hydrodynamic and time-dependent atmosphere models interpreting
   simultaneously both the photometric and spectroscopic observations.
}{
The bolometric light curve of SN 1999em and the spectral evolution of its H$\alpha$
   line are consistent with a presupernova radius of $500\pm200 R_{\sun}$, an ejecta
   mass of $19.0\pm1.2 M_{\sun}$, an explosion energy of $(1.3\pm0.1)\times10^{51}$ erg,
   and a radioactive $^{56}$Ni mass of $0.036\pm0.009 M_{\sun}$.
A mutual mixing of hydrogen-rich and helium-rich matter in the inner layers of
   the ejecta guarantees a good fit of the calculated light curve to that
   observed.
Based on the hydrodynamic models in the vicinity of the optimal model, we derive
   the approximate relationships between the basic physical and observed
   parameters.
The hydrodynamic and atmosphere models of SN 1999em are inconsistent with the short
   distance of 7.85 Mpc to the host galaxy.
}{
We find that the hydrogen recombination in the atmosphere of a normal Type
   IIP SN 1999em, as well as most likely other Type IIP supernovae at the
   photospheric epoch, is essentially a time-dependent phenomenon.
It is also shown that in normal Type IIP supernovae the homologous expansion of
   the ejecta in its atmosphere takes place starting from nearly the third day after
   the supernova explosion.
A comparison of SN 1999em with SN 1987A reveals two very important results for
   supernova theory.
First, the comparability of the helium core masses and the explosion energies
   implies a unique explosion mechanism for these core collapse supernovae.
Second, the optimal model for SN 1999em is characterized by a weaker $^{56}$Ni
   mixing up to $\approx 660$ km\,s$^{-1}$ compared to a moderate $^{56}$Ni
   mixing up to $\sim 3000$ km\,s$^{-1}$ in SN 1987A, hydrogen being mixed
   deeply downward to $\sim 650$ km\,s$^{-1}$.
}
%
\titlerunning{Type IIP supernova 1999em}
\authorrunning{V. P. Utrobin}
\maketitle

\section{Introduction}
The supernova (SN) 1999em was discovered by the Lick Observatory Supernova Search
   on October 29.44 UT in the nearly face-on SBc galaxy NGC 1637
   (Li \cite{li99}).
Detected shortly after the explosion at an unfiltered magnitude of $\sim 13.5$,
   SN 1999em was bright enough to be observed well both photometrically and
   spectroscopically for more than 500 days (Hamuy et al. \cite{hpm01};
   Leonard et al. \cite{lfg02}; Elmhamdi et al. \cite{edc03}).
SN 1999em was the first Type II-plateau supernova (SN IIP) detected at both
   X-ray and radio wavelengths, being the least radio luminous and one of
   the least X-ray luminous SNe (Pooley et al. \cite{plf02}).
The X-ray data indicated the interaction between SN ejecta and
   a pre-SN wind with a low mass-loss rate of $\sim 2 \times 10^{-6}
   M_{\sun}\,yr^{-1}$.
Leonard et al. (\cite{lfab01}) presented the first spectropolarimetry of SN IIP
   based on the optical observations of SN 1999em during $\sim 160$ days after
   SN discovery.
The weak continuum polarization increasing from $p \approx 0.2\%$ on day 7 to
   $p \approx 0.5\%$ in the final observations was detected with an unchanging
   polarization angle $\theta \approx 160^{\circ}$.
To date this event has become the most studied normal SN IIP.

In order to study properly any individual object, an accurate distance is
   required in addition to high quality observations.
It is this point that is not clear for SN 1999em.
There was a reasonable agreement in determining the distance to SN 1999em using
   the expanding photosphere method (EPM; Kirshner \& Kwan \cite{kk74}):
   $7.5 \pm 0.5$ Mpc (Hamuy et al. \cite{hpm01}),
   $8.2 \pm 0.6$ Mpc (Leonard et al. \cite{lfg02}), and
   $7.838 \pm 0.331$ Mpc (Elmhamdi et al. \cite{edc03}).
However, Leonard et al. (\cite{lknt03}) identified 41 Cepheid variable stars in
   the galaxy NGC 1637 and found that the Cepheid distance to the host galaxy
   was $11.7 \pm 1.0$ Mpc.
This distance to SN 1999em is $\sim 50\%$ larger than the earlier EPM distance
   estimates.
Assuming a correlation between the total $^{56}$Ni mass and the explosion energy
   for SNe IIP, Nadyozhin (\cite{n03}) evaluated the distance to the galaxy
   NGC 1637 of 11.08 Mpc close to the Cepheid distance.
Using the spectral-fitting expanding atmosphere model (SEAM), Baron et al.
   (\cite{bnbh04}) obtained the distance to SN 1999em of $12.5 \pm 1.8$ Mpc in
   a good agreement with the Cepheid distance scale.
Finally, studying various ingredients entering the original EPM, Dessart \&
   Hillier (\cite{dh06}) improved this method and also achieved better
   agreement to the Cepheid distance with an estimate of $11.5 \pm 1.0$ Mpc.

Starting the investigation of SN 1999em, a normal SN IIP, it is impossible
   to pass by the well-known and studied SN 1987A in the Large Magellanic Cloud
   (LMC), a peculiar SN IIP.
Recent progress in reproducing the bolometric light curve observed in SN 1987A
   with a modern hydrodynamic model (Utrobin \cite{utr04}) and in modelling the
   H$\alpha$ profile and the Ba II 6142 \AA\ line in SN 1987A at the
   photospheric phase with a time-dependent approach for the SN
   atmosphere (Utrobin \& Chugai \cite{uc02}, \cite{uc05}) makes this
   experience very instructive for other SNe IIP.
First, it turned out that a good agreement between the hydrodynamic models
   and the photometric observations of SN 1987A did not guarantee a correct
   description of this phenomenon as a whole.
Second, the strength of the H$\alpha$ line and its profile provided hard
   constraints on the hydrodynamic models.
The most important lesson from this study of SN 1987A is that we have to take
   \emph{both} the photometric \emph{and} spectroscopic observations into
   account to obtain an adequate hydrodynamic model (Utrobin \cite{utr05}).
Now such an approach should be applied to SN 1999em.

Here we present the comprehensive hydrodynamic study of SN 1999em complemented by
   the atmosphere model with the time-dependent kinetics and energy balance.
A brief description of the hydrodynamic model and the atmosphere model based on
   it is given in Sect.~\ref{sec:snmodel}.
The study of SN 1999em we begin with a construction of the optimal hydrodynamic
   model (Sect.~\ref{sec:optmod}).
Then we continue with a question: at what distance to SN 1999em, short or long,
   the hydrodynamic model and the proper atmosphere model are consistent with
   the photometric and H$\alpha$ line observations (Sect.~\ref{sec:evidis}).
The time development of the optimal hydrodynamic model is presented in
   Sect.~\ref{sec:devmod}, and its general regularities in
   Sect.~\ref{sec:genpro}.
The basic relationships between the physical and observed parameters for
   SNe IIP similar to the SN 1999em event are obtained in Sect.~\ref{sec:phyobs},
   while in Sect.~\ref{sec:sn87a} we address the comparison of SN 1999em with
   SN 1987A.
In Sect.~\ref{sec:disc} we discuss our results from the theoretical and
   observational points of view.
Finally in Sect.~\ref{sec:concl} we summarize the results obtained.

We adopt here a distance of 11.7 Mpc (Leonard et al. \cite{lknt03}),
   a recession velocity to SN 1999em of 800 km\,s$^{-1}$ (Leonard et al.
   \cite{lfg02}), an explosion date of JD 2451476.77, and a total extinction
   $A_V=0.31$ (Baron et al. \cite{bbh00}; Hamuy et al. \cite{hpm01};
   Elmhamdi et al. \cite{edc03}).

\section{Supernova modelling and input physics}
\label{sec:snmodel}
Keeping in mind the importance of a hydrodynamic study and of atmosphere
   modelling, including the time-dependent kinetics and energy balance for the
   interpretation of the SN~1987A phenomenon, we use this approach to
   investigate SN 1999em.

A hydrodynamic model is computed in terms of radiation hydrodynamics in the
   one-group approximation taking into account non-LTE effects on the average
   opacities and the thermal emissivity, effects of nonthermal ionization, and
   a contribution of lines to opacity as in the case of SN 1987A
   (Utrobin \cite{utr04}).
Note that the bolometric luminosity of an SN is calculated by including
   the retardation and limb-darkening effects.

The atmosphere model includes the time-dependent ionization and excitation
   kinetics of hydrogen and other elements, the time-dependent kinetics of
   molecular hydrogen, and the time-dependent energy balance
   (Utrobin \& Chugai \cite{uc05}).
The density distribution, chemical composition, radius of the photosphere,
   and effective temperature are provided by the corresponding hydrodynamic
   model.
The obtained time-dependent structure of the atmosphere is then used to
   calculate synthetic spectra at selected epochs.
The spectra are modelled by the Monte Carlo technique suggesting that the
   photosphere diffusively reflects the incident photons and that the line
   scattering is generally non-conservative and is described in terms of
   the line scattering albedo.
The Thomson scattering on free electrons and Rayleigh scattering on neutral
   hydrogen are also taken into account.

\section{Optimal hydrodynamic model}
\label{sec:optmod}
Elmhamdi et al. (\cite{edc03}) have constructed the ``UBVRI'' bolometric light
   curve of SN 1999em from the corresponding photometric data.
To account for a possible contribution of the missing infrared bands, we add
   a value of 0.19 dex, taken from Elmhamdi et al. (\cite{edc03}), to the
   ``UBVRI'' luminosity and adopt the resultant light curve as the bolometric
   light curve of SN 1999em.
Our aim is to find an adequate hydrodynamic model that reproduces the photometric
   and spectroscopic observations of SN 1999em.
To fit the bolometric light curve of SN 1999em, various hydrodynamic models
   were explored.
The bolometric light curve of SN 1999em is fitted by adjusting the pre-SN
   radius $R_0$, the ejecta mass $M_{env}$, and the explosion energy $E$, along
   with the density distribution in the pre-SN model and its chemical
   composition in the transition region between the hydrogen-rich envelope and
   metal/helium core.
The radioactive $^{56}$Ni mass is reliably measured by the light curve tail
   after the plateau phase.
As the pre-SN model of SN 1999em, we consider non-evolutionary models similar
   to those of SN 1987A (Utrobin \cite{utr04}), but for the outer layers assume
   the standard solar composition with the mass fraction of hydrogen $X=0.735$,
   helium $Y=0.248$, and the metallicity $Z=0.017$ (Grevesse \& Sauval
   \cite{gs98}) taking a normal spiral nature of the host galaxy NGC 1637
   into account.
The best version of such a fitting was obtained with the optimal model
   that is characterized by the basic parameters: the pre-SN radius of
   500 $R_{\sun}$, the ejecta mass of 19 $M_{\sun}$, and the explosion energy
   of $1.3\times10^{51}$ erg (model D11 in Table~\ref{tab:hydmods}).
\begin{figure}[t]
   \resizebox{\hsize}{!}{\includegraphics{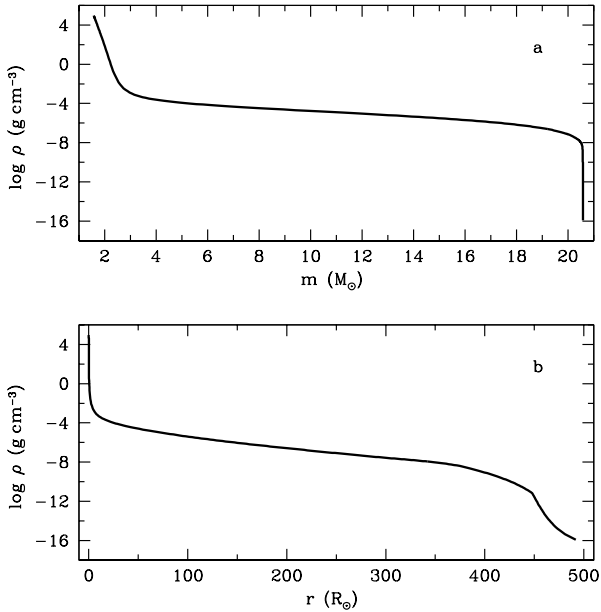}}
   \caption{%
   Density distribution with respect to interior mass (\textbf{a}) and
   radius (\textbf{b}) for the pre-SN model D11.
   The central core of 1.58 $M_{\sun}$ is omitted.
   }
   \label{fig:rhomr}
\end{figure}
\begin{figure}[t]
   \resizebox{\hsize}{!}{\includegraphics[clip, trim=0 0 0 184]{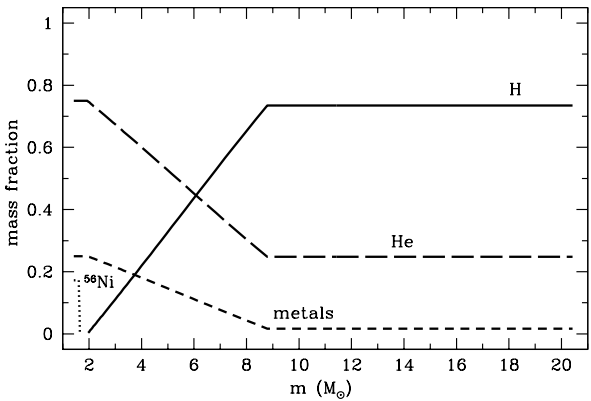}}
   \caption{%
   The mass fraction of hydrogen (\emph{solid line\/}), helium
      (\emph{long dashed line\/}), heavy elements (\emph{short dashed line\/}),
      and radioactive $^{56}$Ni (\emph{dotted line\/}) in the ejecta of
      model D11.
   }
   \label{fig:chcom}
\end{figure}

The density profile of the pre-SN model consisting of a central white
   dwarf like core and an outer envelope with the size of the red supergiant
   is shown in Fig.~\ref{fig:rhomr}.
In the calculations, the 1.58 $M_{\sun}$ central core is removed from the
   computational mass grid and assumed to collapse to a neutron star,
   while the rest of the star is ejected by the SN explosion, which is
   modelled by a piston action near the edge of the central core.
The pre-SN model has a heterogeneous chemical composition containing
   a 5.6 $M_{\sun}$ helium core and a 11.9 $M_{\sun}$ outer shell
   of the solar chemical composition (Fig.~\ref{fig:chcom}).
Note that the helium cores up to about 8 $M_{\sun}$ are consistent with the
   observations (Sect.~\ref{sec:grg-chcom}).
There is no sharp boundary between the hydrogen-rich and helium-rich layers
   in the ejecta of the optimal model.
Hydrogen-rich material is mixed into the central region, and helium-rich
    material, in turn, is mixed outwards.
It is evident that such a distribution of hydrogen and helium implies a strong
   mixing at the helium/hydrogen composition interface.
The fact that the radioactive $^{56}$Ni is confined to the innermost ejected
   layers suggests its weak mixing during the SN explosion.
\begin{figure}[t]
   \resizebox{\hsize}{!}{\includegraphics[clip, trim=0 0 0 184]{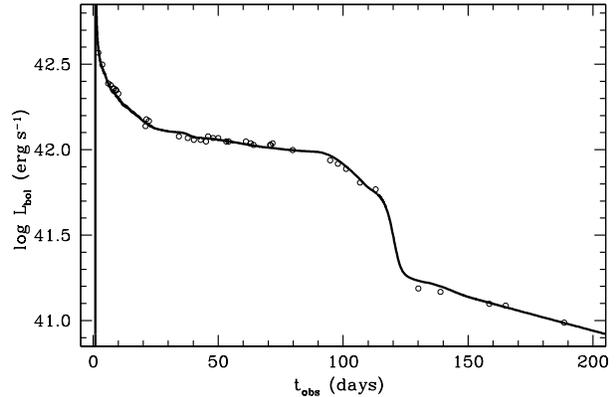}}
   \caption{%
   Comparison of the calculated bolometric light curve of model D11
      (\emph{solid line\/}) with the bolometric data of SN 1999em
      obtained by Elmhamdi et al. (\cite{edc03}) (\emph{open circles\/}).
   }
   \label{fig:lmbol}
\end{figure}

In Fig.~\ref{fig:lmbol} we show the very good match between the bolometric
   light curve calculated for the optimal hydrodynamic model and the one observed
   for SN 1999em (Elmhamdi et al. \cite{edc03}).
Note that hereafter $t_{obs}$ is the time in the observer's frame of reference.
The model agrees well with the observed tail of the bolometric light curve
   for the total $^{56}$Ni mass of 0.036 $M_{\sun}$, the bulk of the
   radioactive $^{56}$Ni being mixed in the velocity range $\le 660$ km\,s$^{-1}$
   (Fig.~\ref{fig:denicl}) in order to reproduce the observed transition from
   the plateau to the tail.

\section{H{\boldmath $\alpha$} profile: evidence against short distance}
\label{sec:evidis}
%
\begin{figure}[t]
   \resizebox{\hsize}{!}{\includegraphics{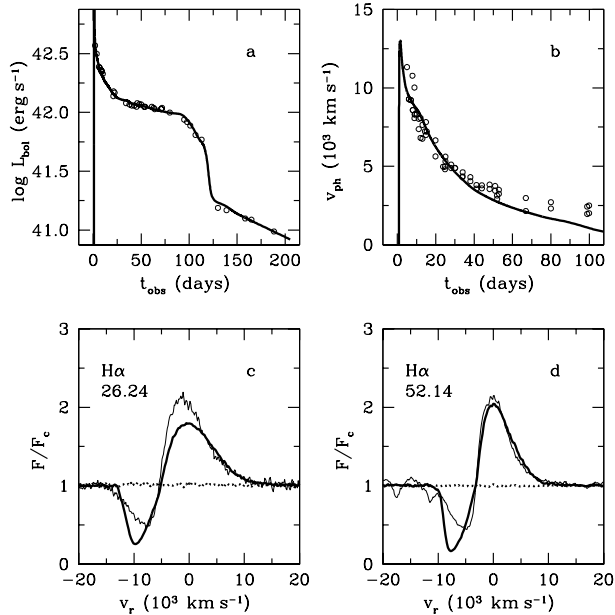}}
   \caption{%
   Hydrodynamic model D11 and H$\alpha$ line for the distance of 11.7 Mpc.
   Panel (\textbf{a}): the calculated bolometric light curve (\emph{solid
      line\/}) compared with the observations of SN 1999em obtained by
      Elmhamdi et al. (\cite{edc03}) (\emph{open circles\/}).
   Panel (\textbf{b}): calculated photospheric velocity (\emph{solid
      line\/}) and radial velocities at maximum absorption of spectral lines
      measured by Hamuy et al. (\cite{hpm01}) and Leonard et al. (\cite{lfg02})
      (\emph{open circles\/}).
   Panel (\textbf{c}): H$\alpha$ profiles, computed with the time-dependent
      approach (\emph{thick solid line\/}) and with the steady-state model
      (\emph{dotted line\/}), overplotted on the observed profile on
      day 26.24, as obtained by Leonard et al. (\cite{lfg02}) (\emph{thin solid
      line\/}).
   Panel (\textbf{d}): the same as panel (c) but for day 52.14.
   }
   \label{fig:hmd117}
\end{figure}
\begin{figure}[t]
   \resizebox{\hsize}{!}{\includegraphics{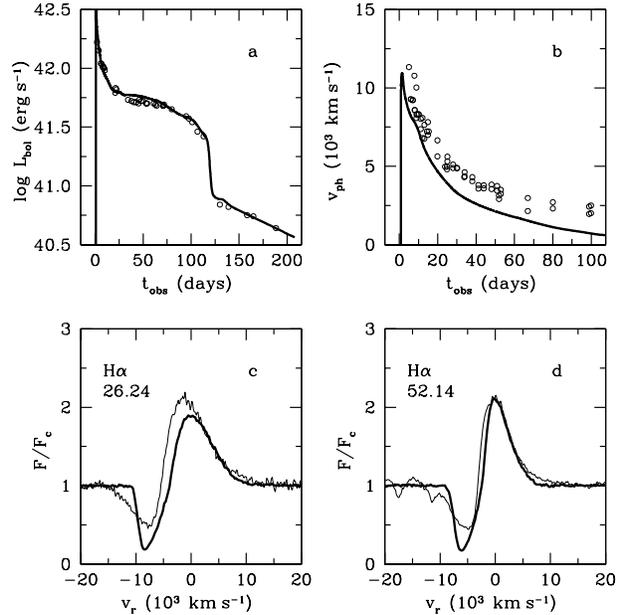}}
   \caption{%
   Hydrodynamic model D07 and H$\alpha$ line for the distance of 7.85 Mpc.
   See the Fig.~\ref{fig:hmd117} legend for details.
   }
   \label{fig:hmd785}
\end{figure}
It is quite clear that any well-observed SN should be described by a unique
   hydrodynamic model in combination with the atmosphere model based on it.
The experience during the study of SN 1987A showed that such
   a combination of the models had to fit not only the photometric, but
   also the spectroscopic observations.
We believe that the adequate hydrodynamic and atmosphere models of SN 1999em
   are able to distinguish between the short and long distances.
A distance of 7.85 Mpc, the average value of the EPM distance estimates,
   is taken as the short distance, and the Cepheid distance of 11.7 Mpc is
   taken as the long distance.
\begin{table}[t]
\caption[]{Hydrodynamic models for distances of 7.85 and 11.7 Mpc.}
\label{tab:hydmods}
\centering
\begin{tabular}{c  c  c @{ } c @{ }  c  c  c}
\hline\hline
\noalign{\smallskip}
 Model & $R_0$ & $M_{env}$ & $E$ & $M_{\mathrm{Ni}}$ & $X$ & $Z$ \\
       & $(R_{\sun})$ & $(M_{\sun})$ & ($10^{51}$ erg) & $(10^{-2} M_{\sun})$ &
       & \\
\noalign{\smallskip}
\hline
\noalign{\smallskip}
D07 & 375 & 16 & 0.686 & 1.62 & 0.735 & 0.017 \\
D11 & 500 & 19 & 1.300 & 3.60 & 0.735 & 0.017 \\
\noalign{\smallskip}
\hline
\end{tabular}
\end{table}

As demonstrated above, we constructed the optimal hydrodynamic model for the
   long distance (model D11 in Table~\ref{tab:hydmods}).
The model reproduces the observed bolometric light curve of SN 1999em very well
   (Fig.~\ref{fig:hmd117}a).
In Fig.~\ref{fig:hmd117}b the calculated expansion velocity at the photosphere
   level, the photospheric velocity, is compared with the radial velocities
   at maximum absorption of the different spectral lines measured by Hamuy et al.
   (\cite{hpm01}) and Leonard et al. (\cite{lfg02}).
The photospheric velocity of model D11 is consistent with the observed points,
   at least for the first 60 days.
To verify the hydrodynamic model by matching the constraints from the spectral
   observations of SN 1999em, we examined the H$\alpha$ profile on days 26.24
   and 52.14 computed in the time-dependent approach.

In the time-dependent atmosphere model for SN 1987A, we considered two extreme
   cases to allow for the uncertainty of our approximation in a description of
   the ultraviolet radiation field: (i) the photospheric brightness is black-body
   with the effective temperature (model A); (ii) the photospheric brightness
   corresponds to the observed spectrum (model B) (Utrobin \& Chugai
   \cite{uc05}).
For SN 1999em the photospheric brightness in model B is black-body with the
   effective temperature and the corresponding brightness reduction taken from
   the SN 1987A observations.

The time-dependent approach with model B satisfactorily reproduces the strength
   of the H$\alpha$ emission component on day 26.24 with some emission deficit
   near the maximum in comparison to what was observed (Fig.~\ref{fig:hmd117}c).
This deficit is not significant because model A, calculated for this phase and
   not plotted on Fig.~\ref{fig:hmd117}c for the sake of clarity, gives
   the emission component with a relative flux of 3.1 at the maximum, which
   is much higher than the observed one, so the real situation is somewhere
   between these two cases and is closer to model B.
Note that in SN 1987A the H$\alpha$ emission component demonstrated the same
   behavior in the early phase (Utrobin \& Chugai \cite{uc05}).
In contrast, the H$\alpha$ absorption component calculated in model B is
   stronger than that observed on day 26.24 in SN 1999em.
This discrepancy is presumably related to a poor description of the ultraviolet
   radiation at frequencies between the Balmer and Lyman edges.
This radiation interacts with numerous metal lines and controls the populations
   of hydrogen levels.
It is very important that both the emission and absorption components of the
   H$\alpha$ line calculated in model B extend over the whole range of the
   observed radial velocities from $-15\,000$ km\,s$^{-1}$ to $15\,000$
   km\,s$^{-1}$ (Fig.~\ref{fig:hmd117}c).
Unfortunately, the calculated absorption component runs above the observed one
   in a radial velocity range between $-15\,000$ km\,s$^{-1}$ and $-12\,500$
   km\,s$^{-1}$.
On day 52.14 the emission component of the H$\alpha$ line computed with the
   time-dependent approach with model B fits the observed one fairly well,
   while the absorption component is still stronger than the observed one
   (Fig.~\ref{fig:hmd117}d).
Thus, we may state that the above hydrodynamic and atmosphere models are in
   a good agreement with the photometric and spectroscopic observations of
   SN 1999em.

Now let us pay attention to the time-dependent effects in hydrogen lines of
   SN 1999em, a normal SN~IIP.
The time-dependent approach with model B reproduces the strength of the H$\alpha$
   line at day 26.24 and day 52.14 fairly well (Figs.~\ref{fig:hmd117}c and
   \ref{fig:hmd117}d).
In contrast, a steady-state model B demonstrates an extremely weak H$\alpha$
   line on days 26.24 and 52.14.
This reflects the fact that the steady-state ionization is significantly
   lower than in the time-dependent model, while the electron temperature
   is too low for the collisional excitation of hydrogen.
The radioactive $^{56}$Ni is mixed too weakly to affect the ionization and
   excitation of hydrogen and other elements in the atmosphere at the plateau
   phase.
Thus, it is possible to conclude that the hydrogen recombination in the
   atmosphere of SN 1999em during the whole plateau phase is essentially
   a time-dependent phenomenon.

In turn, the short distance results in the hydrodynamic model with the
   following parameters: the pre-SN radius of 375 $R_{\sun}$,
   the ejecta mass of 16 $M_{\sun}$,  and the explosion energy of
   $6.86\times10^{50}$ erg (model D07 in Table~\ref{tab:hydmods}).
In this model the central core of 1.4 $M_{\sun}$ is assumed to collapse
   to a neutron star.
The hydrodynamic model fits the observed bolometric light curve
   of SN 1999em but for the $^{56}$Ni mass of 0.0162 $M_{\sun}$, the most of
   $^{56}$Ni being mixed in the velocity range $\le 580$ km\,s$^{-1}$
   (Fig.~\ref{fig:hmd785}a).
The photospheric velocity curve runs well below the observed points
   (Fig.~\ref{fig:hmd785}b), and it might be expected that the spectral lines
   in this model would be narrower than those observed.
Indeed, on days 26.24 and 52.14 both the emission and absorption components
   of the H$\alpha$ line computed with the time-dependent approach with model B
   are significantly narrower than those observed in SN 1999em
   (Figs.~\ref{fig:hmd785}c and \ref{fig:hmd785}d).
The hydrodynamic model D07 agrees fairly well with the observed bolometric
   light curve, but the relevant atmosphere model fails to reproduce the
   H$\alpha$ profile observed in SN 1999em.
We thus conclude that the short distance of 7.85 Mpc should be discarded.

\section{Development of the optimal model}
\label{sec:devmod}
%
\begin{figure}[t]
   \resizebox{\hsize}{!}{\includegraphics[clip, trim=0 0 0 184]{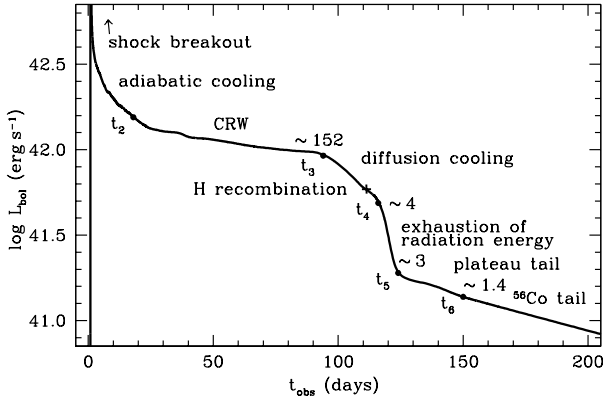}}
   \caption{%
   Evolution of a normal SN~IIP illustrated by the bolometric light curve
      of the optimal model.
   Lettered dots mark the specific times in the evolution.
   The end time of the shock breakout phase, $t_1$, is shown in
      Fig.~\ref{fig:swbrck}a.
   The cross notes the moment of complete hydrogen recombination.
   The numbers indicate the total optical depth of the ejecta at the
      corresponding times.
   }
   \label{fig:snphs}
\end{figure}
Although the major issues of light curve theory were recognized a long time
   ago (Grassberg et al. \cite{gin71}; Falk \& Arnett \cite{fa77}), it is useful
   to consider an SN outburst in a more detailed approach.
This study of SN 1999em provides a good opportunity to examine the time
   development of a normal SN~IIP.
Figures~\ref{fig:snphs} and \ref{fig:swbrck}a show the following stages
   in the observer time scale: a shock breakout ($t_{obs} \le t_1$),
   an adiabatic cooling phase ($t_1 < t_{obs} \le t_2$),
   a phase of cooling and recombination wave (CRW) ($t_2 < t_{obs} \le t_3$),
   a phase of radiative diffusion cooling ($t_3 < t_{obs} \le t_4$),
   an exhaustion of radiation energy ($t_4 < t_{obs} \le t_5$),
   a plateau tail ($t_5 < t_{obs} \le t_6$), and a radioactive tail
   ($t_{obs} > t_6$).
In addition to the above stages there are two specific points: a transition
   from acceleration of the envelope matter to a homologous expansion
   and a moment of complete hydrogen recombination.
In the optimal model D11, the characteristic moments are $t_1 \approx 0.93$ days,
   $t_2 \approx 18$ days, $t_3 \approx 94$ days, $t_4 \approx 116$ days,
   $t_5 \approx 124$ days, and $t_6 \approx 150$ days, and the complete hydrogen
   recombination occurs at $t_\mathrm{H} \approx 111.3$ days.
Below we consider the basic stages in the time development of the optimal model.

\subsection{Shock breakout}
\label{sec:dmd-shbrk}
%
\begin{figure}[t]
   \resizebox{\hsize}{!}{\includegraphics{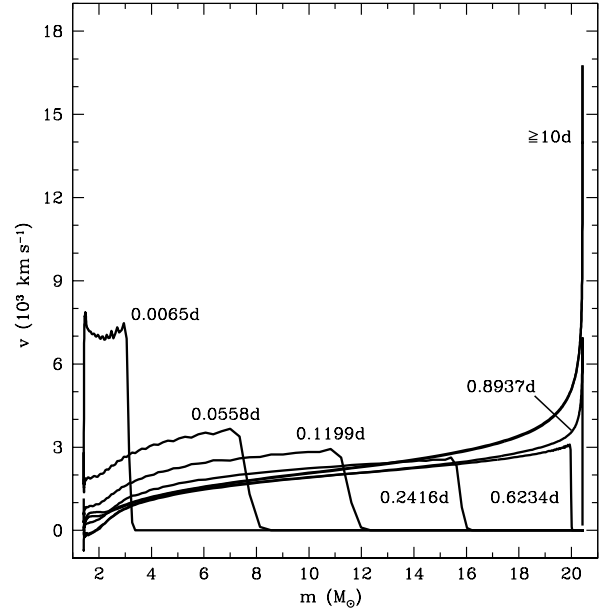}}
   \caption{%
   Propagation of the shock through the ejected envelope of model D11.
   Velocity profiles with respect to interior mass are plotted at
      $t=0.0065$, 0.0558, 0.1199, 0.2416, 0.6234, 0.8937 days, and
      $t \ge 10$ days after the SN explosion.
   }
   \label{fig:velvsm}
\end{figure}
The explosion of the star is assumed to be triggered by a piston action
   near the edge of the central core immediately after the epoch zero,
   $t=0$.
From here on, $t$ is the time in the comoving frame of reference.
This energy release generates a strong shock wave that propagates towards
   the stellar surface.
In moving out of the center, the shock wave heats matter and accelerates it
   to velocities increasing outward and exceeding the local escape velocity.
From $t=0.0065$ days to $t=0.2416$ days, the shock wave, propagating outward
   from the compact dense core of the pre-SN (Fig.~\ref{fig:rhomr}),
   is attenuated slightly due to the spherical divergence
   (Fig.~\ref{fig:velvsm}).
It then reaches the outermost layers with a sharp decline in density and after
   $t=0.6234$ days gains strength and accelerates due to the effect of
   hydrodynamic cumulation.
Only a small portion of the star undergoes this acceleration and acquires
   a high velocity.
For example, the layers of velocities exceeding 5000~km~s$^{-1}$ have a mass
   of $\approx 0.478 M_{\sun}$.
\begin{figure}[t]
   \resizebox{\hsize}{!}{\includegraphics{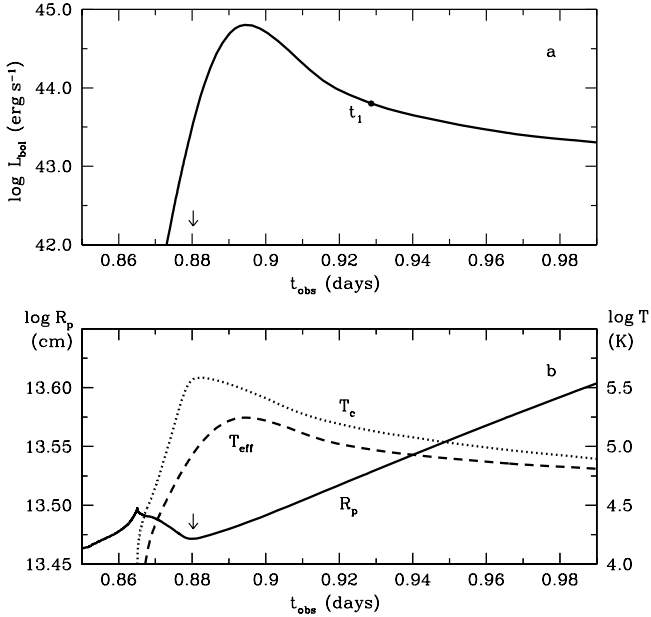}}
   \caption{%
   Shock breakout in model D11. Arrow indicates the moment of 0.8803 days
      when the velocity at the stellar surface reaches escape velocity.
   Panel (\textbf{a}): the calculated bolometric light curve and the lettered
      dot marking the end time of the shock breakout phase.
   Panel (\textbf{b}): the calculated photospheric radius (\emph{solid line\/}),
      the effective temperature (\emph{dashed line\/}), and the color
      temperature (\emph{dotted line\/}) as a function of the observer's time.
   }
   \label{fig:swbrck}
\end{figure}

By day 0.8651 the shock wave reaches the stellar surface and then begins to heat
   the external layers, so that the color temperature jumps to
   $3.84 \times 10^{5}$ K at day 0.8823, the effective temperature increases up
   to $1.76 \times 10^{5}$ K at day 0.8943, and the bolometric luminosity rises
   accordingly up to $6.46 \times 10^{44}$ erg\,s$^{-1}$ at day 0.8943
   (Fig.~\ref{fig:swbrck}).
Note that the maximum of the color temperature coincides closely with the moment
   of 0.8803 days when a velocity at the stellar surface reaches the escape
   velocity.

A very rapid rise in the bolometric luminosity to maximum, starting at
   day 0.8651, instantly changes a growth of the photospheric radius into
   its reduction because of the intense radiative losses of energy in the
   outermost layers (Fig.~\ref{fig:swbrck}b).
At day 0.8691 these layers begin to move outward, and the additional cooling
   by adiabatic expansion makes the reduction of the photospheric radius more
   noticeable.
At the same time the envelope expansion is favorable to a photon diffusion,
   decreasing the characteristic diffusion time.
The photon diffusion eventually dominates, stops this reduction at day
   0.8802, and then, along with the envelope expansion, blows the photospheric
   radius outward.

When a velocity at the stellar surface exceeds the escape velocity, the outside
   layers of the star begin to cool rapidly, the color temperature begins to
   decrease from its maximum value, and both the effective temperature and
   the luminosity decrease somewhat later.
A narrow peak in the bolometric luminosity forms as a result
   (Fig.~\ref{fig:swbrck}a).
The peak has a width of about $0.02$ days at a half level of the luminosity
   maximum.
Most of its radiation is emitted in an ultraviolet flash.
The total number of ionizing photons above 13.598 eV for the whole outburst is
   $2.768 \times 10^{58}$.
During the first 1.234 days the number of ionizing photons is $90\%$ of the
   total number, and the radiated energy adds up to $1.59 \times 10^{48}$ ergs.
We conditionally define the transition time between the shock breakout and
   the adiabatic cooling phase, $t_1$, as the time at which the bolometric
   luminosity drops by one dex from its maximum value, with the understanding
   that the adiabatic cooling becomes essential soon after the onset of
   the envelope expansion.
In the optimal model this transition time is nearly 0.93 days.

Scattering processes are fundamentally nonlocal and result in exceeding the
   color temperature over the effective one when they are dominant (Mihalas
   \cite{m78}; Sobolev \cite{s80}).
During the shock breakout, the adiabatic cooling phase and the CRW phase
   scattering processes dominate those of true absorption near the
   photospheric level in the ejecta of the optimal model and, as
   a consequence, lead to a color temperature that is considerably higher than
   the effective temperature.
Figure~\ref{fig:swbrck}b shows that the maximum of the color temperature
   coincides very closely in time with the local minimum of the photospheric
   radius at day 0.8802 and comes before the maximum of the effective
   temperature and the bolometric luminosity.

\subsection{Adiabatic cooling phase}
\label{sec:dmd-adbcln}
During the shock passage throughout the star, the gas is heated up to about
   $10^{5}$ K and, as a consequence, is totally ionized.
Both the internal gas energy and the radiation energy increase in the SN
   envelope.
After the shock breakout, a dominant process in the subphotospheric, optically
   thick layers of the expelled envelope is the cooling by adiabatic expansion.
The adiabatic losses drastically reduce the stored energy and completely
   determine the evolution of the bolometric luminosity during the adiabatic cooling
   phase (Figs.~\ref{fig:snphs} and \ref{fig:swbrck}a).
Such behavior of the luminosity lasts till the gas and radiation temperatures
   drop to the critical value in the subphotospheric layers, and a recombination
   of hydrogen, the most abundant element in the ejected envelope, starts.
In these layers hydrogen becomes partially ionized by around day 18.
For the adiabatic cooling phase, the color and effective temperatures drop
   rapidly from $1.35 \times 10^{5}$ K and $9.32 \times 10^{4}$ K, respectively,
   at day 0.93 to 6650 K and 6560 K at day 18.
Their ratio also reduces as the contribution of scattering processes to the
   opacity decreases.

In an optically thick medium, the strong shock wave propagates almost adiabatically
   throughout the stellar matter.
When the shock wave emerges on the pre-SN surface, the adiabatic regime
   is broken and transforms into the isothermal regime.
This transformation takes place during the adiabatic cooling phase and gives
   rise to a thin, dense shell at the outer edge of the ejected envelope.
The dense shell is arising from day 0.945 to day 1.022, reaching the density
   contrast of $\sim$ 210 at a velocity of $\sim$ 12\,300 km\,c$^{-1}$.
The formation of the dense shell starts in the optically thick medium at the
   optical depth of $\sim$ 8, and ends in semi-transparent medium at the
   optical depth of $\sim$ 1.
By day 18 the shell accelerates to a velocity of $\sim$ 13\,400 km\,c$^{-1}$
   reducing the density contrast to a value of $\sim$ 2.5.
The matter in the dense shell is subject to the Rayleigh-Taylor instability,
   which may result in the strong mixing of matter (Falk \& Arnett \cite{fa77}).
The latter, in turn, may prevent the thin shell-like structure from forming
   in the outer layers of the envelope.

\subsection{Homologous expansion}
\label{sec:dmd-free}
%
\begin{figure}[t]
   \resizebox{\hsize}{!}{\includegraphics{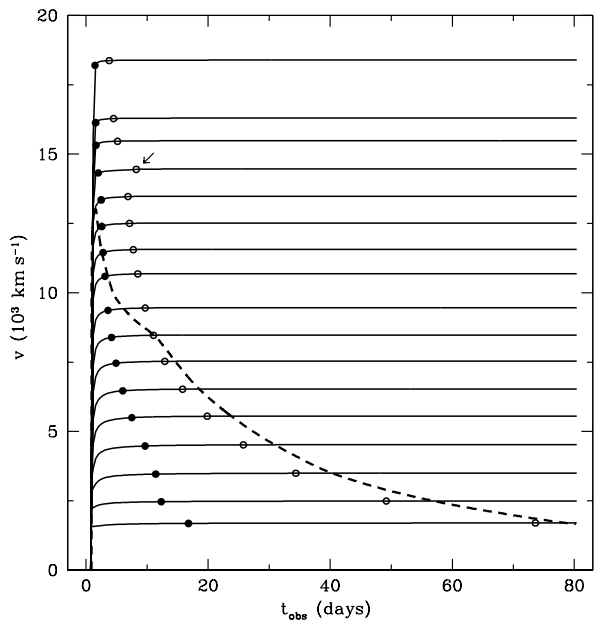}}
   \caption{%
   Time dependence of the velocity of different mass shells (\emph{solid line\/})
      in the ejected envelope of model D11 and the photospheric velocity
      (\emph{dashed line\/}).
   \emph{Full} and \emph{open circles} indicate moments when a velocity of
      the mass shell reaches $99\%$ and $99.9\%$, respectively, of its
      terminal velocity.
   The arrow marks the mass shell that has undergone an additional acceleration
      due to the expansion opacity.
   }
   \label{fig:freexp}
\end{figure}
The SN explosion causes an acceleration of the envelope matter.
At late times, when the acceleration of the ejecta becomes negligible,
   the envelope matter expands homologously.
Evidently, there is a transition from the acceleration to the homologous
   expansion.
This transition for different mass shells occurs at different times:
   the deeper the layer, the later the transition (Fig.~\ref{fig:freexp}).
Thus, the transition time of the whole expelled envelope is determined by the
   deepest layer.
If we consider physical processes in an SN atmosphere, the relevant
   transition time is given by the photospheric location.
It is clear that the transition time for the SN atmosphere is much
   shorter than for the whole envelope.
For instance, the atmosphere layers exceed $99\%$ and $99.9\%$ of their terminal
   velocities starting from nearly day 2.8 and day 11, respectively
   (Fig.~\ref{fig:freexp}).

It is interesting that a monotonic increase in the transition time with
   embedding inward the ejecta is broken at the $99.9\%$ level of the terminal
   velocity.
There is a prominent feature at a velocity of $\sim$ 14\,500 km\,c$^{-1}$ marked
   in Fig.~\ref{fig:freexp}.
As we show later in Sect.~\ref{sec:grg-lop}, it is the result of
   an additional acceleration due to the resonance scattering in numerous metal
   lines.
\begin{figure}[t]
   \resizebox{\hsize}{!}{\includegraphics{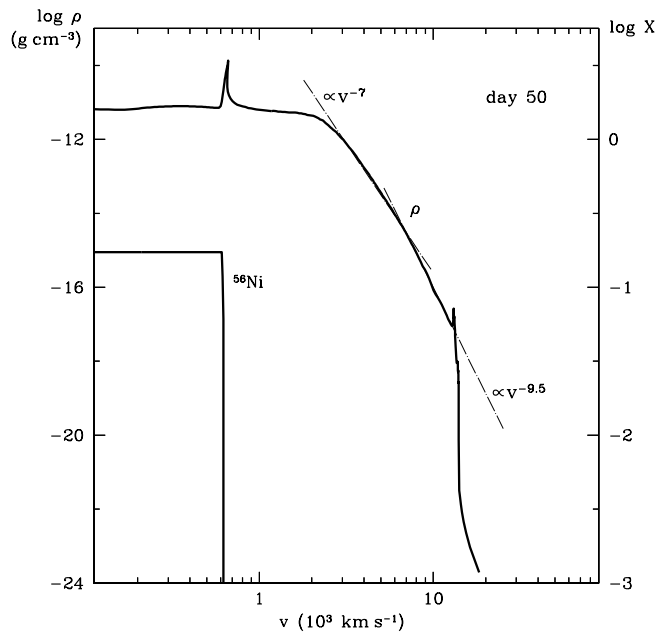}}
   \caption{%
   The density and the $^{56}$Ni mass fraction as a function of the velocity
      for model D11 at $t=50$ days.
   \emph{Dot-dash lines} are the density distribution fits $\rho \propto v^{-7}$
   and $\rho \propto v^{-9.5}$.
   }
   \label{fig:denicl}
\end{figure}

Homologous expansion is characterized by a density distribution frozen in
   velocity space and scaled in time as $\propto t^{-3}$.
Such a density profile for the optimal model is shown at $t=50$ days in
   Fig.~\ref{fig:denicl}.
In the inner region of the ejected envelope there is a dense shell with
   a density contrast of $\sim$ 20 produced by the $^{56}$Ni bubble at
   a velocity of $\sim$ 660 km\,c$^{-1}$.
The density distribution above the $^{56}$Ni bubble shell is nearly uniform
   up to $\sim$ 2000 km\,c$^{-1}$.
The outer layers with velocities in the ranges 3000--6900~km~s$^{-1}$ and
   6900--13\,000~km~s$^{-1}$ may be fitted by an effective index
   $n=-\partial \ln \rho / \partial \ln r$ of 7 and 9.5, respectively, as seen
   from the density distribution with respect to the expansion velocity
   (Fig.~\ref{fig:denicl}).
At a velocity of $\sim$ 13\,000 km\,c$^{-1}$, there is another dense shell with
   a density contrast of $\sim$ 3 originated during the adiabatic cooling phase.
Note that both dense shells slightly change their density profiles with
   time.
For example, the evolution of the $^{56}$Ni bubble shell is clearly demonstrated
   by Fig.~\ref{fig:crwdnt}a.

\subsection{Cooling and recombination wave}
\label{sec:dmd-crw}
%
\begin{figure}[t]
   \resizebox{\hsize}{!}{\includegraphics{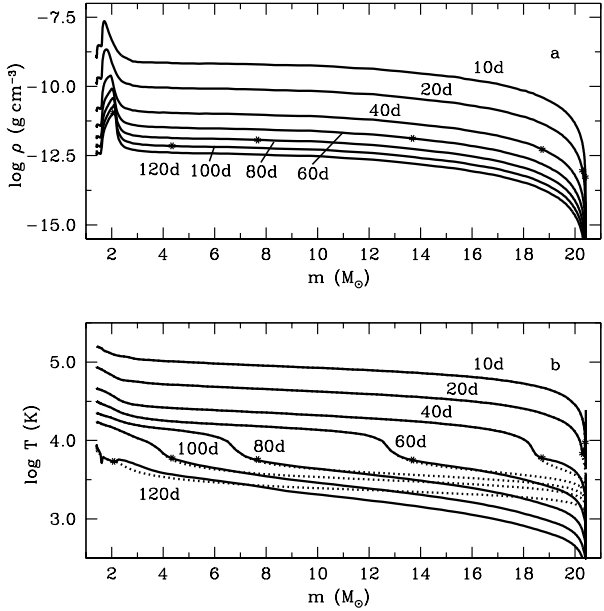}}
   \caption{%
   Evolution of the density (\textbf{a}) and the gas (\emph{solid line\/})
      and radiation (\emph{dotted line\/}) temperatures (\textbf{b}) as
      a function of mass for model D11.
   Profiles are plotted at $t=10$, 20, 40, 60, 80, 100, and 120 days.
   Star indicates the position of the photosphere.
   }
   \label{fig:crwdnt}
\end{figure}
\begin{figure}[t]
   \resizebox{\hsize}{!}{\includegraphics{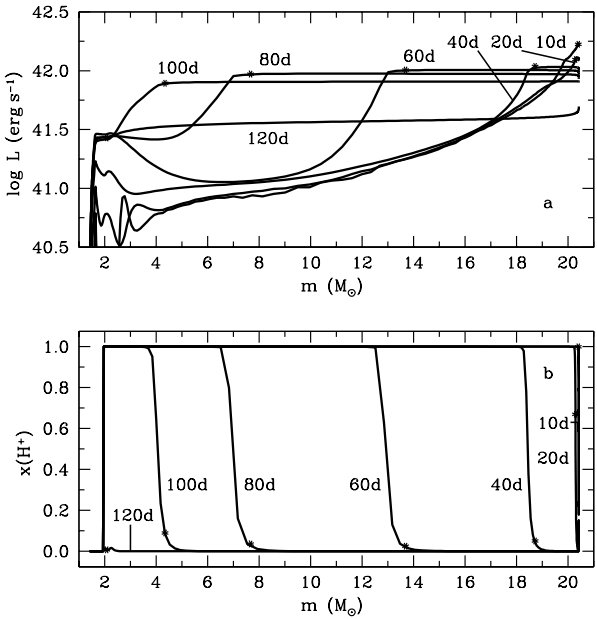}}
   \caption{%
   Evolution of the interior luminosity (\textbf{a}) and the fraction of ionized
      hydrogen (\textbf{b}) as a function of mass for model D11.
   Profiles are plotted at $t=10$, 20, 40, 60, 80, 100, and 120 days.
   Star indicates the position of the photosphere.
  }
   \label{fig:crwlxh}
\end{figure}
As the envelope expands, cooling by radiation -- a cooling and recombination
   wave -- occurs and completely dominates the luminosity of the SN
   by about day 18.
From this time to nearly day 94 the bolometric luminosity is mainly determined
   by properties of the CRW (for details see Grassberg \& Nadyozhin \cite{gn76};
   Imshennik \& Nadyozhin \cite{in89}).
A main property of the CRW consists in generating virtually the entire energy
   flux carried away by radiation within its front.
This radiation flux exhausts the thermal and recombination energy of
   cooling matter by means of recombination.
At the inner edge of the CRW, the radiation flux is negligibly small and matter
   is ionized, but at the outer edge the flux is equal to the luminosity of the
   star, and matter completely recombines.

During the CRW phase the evolution of the density in the ejecta illustrates the
   homologous expansion (Fig.~\ref{fig:crwdnt}a).
The behavior of the gas and radiation temperatures reflects two important facts
  (Fig.~\ref{fig:crwdnt}b).
First, in the subphotospheric layers in an optically thick medium, under
   conditions very close to LTE, the radiation temperature is virtually equal
   to the gas temperature.
Second, in the transparent layers, beyond the photosphere, the gas temperature
   turns out to be lower than the radiation temperature because of the weak
   interaction between matter and radiation and of the dominant role of
   adiabatic losses.
From an inspection of the interior luminosity and the fraction of ionized
   hydrogen from $t=20$ days to $t=100$ days, the CRW characteristics described
   above are evident (Figs.~\ref{fig:crwlxh}a and \ref{fig:crwlxh}b).
Hence in this period the CRW occurs.

From $t=20$ days to $t=100$ days, the CRW propagates through a mass of
   $\approx 15.96 M_{\sun}$, while the outermost layers, those responsible for
   the luminosity during the first 18 days, contribute only about
   $0.13 M_{\sun}$.
It is interesting that for this period the gas and radiation temperatures at
   the photosphere position are nearly constant in time and equal to about
   5600 K (Fig.~\ref{fig:crwdnt}b).
And from $t=40$ days to $t=100$ days, the photosphere is located at the outer
   edge of the hydrogen recombination front (Fig.\ref{fig:crwlxh}b).

After $t=100$ days the radiation flux at the inner edge of the CRW becomes
   comparable to the luminosity of star (Fig.~\ref{fig:crwlxh}a).
This effect is caused by radiative diffusion from the central envelope layers
   where the energy of radioactive decays is being released.
Later on the luminosity is entirely determined by radiative diffusion and not
   by the CRW.

\subsection{Radiative diffusion cooling}
\label{sec:dmd-difcln}
A cooling by radiative diffusion starts in the optically thick expelled envelope
   at day 94 and ends in the semi-transparent medium at day 116
   (Fig.~\ref{fig:snphs}).
Shortly before the end of the phase of radiative diffusion cooling, the complete
   hydrogen recombination occurs at $t_\mathrm{H} \approx 111.3$ days.
In the optimal model the radiative diffusion takes place in the inner layers
   with the nearly uniform density distribution and results in the
   characteristic shoulder of the bolometric luminosity.
It should be noted that this behavior of the light curve was discussed by Falk
   \& Arnett (\cite{fa77}).
To describe photon diffusion, we may use a simple one-zone approximation (Arnett
   \cite{arn79}), ignoring the contribution of radioactive decays to the
   bolometric light curve:
\begin{equation}
   {d \ln L_{bol} \over d t} = - {1 \over t_{dif}} \; , \,
   t_{dif} = {\pi \over 3}{R_d \over c} \tau_d \; ,
   \label{eq:lboldif}
\end{equation}
where $t_{dif}$ is the diffusion time, $R_d$ the characteristic radius of
   the diffusion region, and $\tau_d$ the optical depth of this region.

Evidently, it is possible to estimate the diffusion time from both the decline
   of the bolometric light curve and the physical properties of the diffusion
   region.
The decline of the luminosity shoulder between day 94 and day 116 measures
   a diffusion time of $\approx 35$ days (Fig.~\ref{fig:snphs}).
On the other hand, the radius of the diffusion region of
   $\approx 1.2 \times 10^{15}$ cm and its optical depth of $\approx 63$
   evaluated at day 105 give a diffusion time of $\approx 30$ days.
These estimates are consistent with the above description of photon diffusion
   (\ref{eq:lboldif}).
Thus, the luminosity shoulder at the phase of radiative diffusion cooling is
   approximated well by the simple model with the constant diffusion time.
Note that a transition from the constant diffusion time to the variable
   diffusion time, which is getting shorter, transforms the linear shoulder of
   the luminosity in logarithmic scale into a convex light curve at the end
   of the plateau.

\subsection{Exhaustion of radiation energy}
\label{sec:dmd-exrden}
During the entire outburst, the total radiation energy in the ejecta is much greater
   than its total internal gas energy.
In the inner optically thick layers of the envelope, in which radiation is in
   equilibrium with matter, the radiative diffusion reduces both the radiation
   and internal gas energies.
By day $116$ the total optical depth of the ejecta falls to a value of $\sim 4$,
   and a low opacity of matter significantly weakens the interaction between
   radiation and matter.
As a consequence, for the next 9 days the photon diffusion mainly reduces the
   stored radiation energy and eventually exhausts it, and the luminosity
   decreases abruptly and becomes close to the gamma-ray luminosity due
   to the decay of the radioactive $^{56}$Co, which is calculated by taking
   account of the retardation effect in the envelope transparent for gamma-rays
   (Fig.~\ref{fig:scndpl}a).

\subsection{Plateau tail phase}
\label{sec:dmd-secnd}
%
\begin{figure}[t]
   \resizebox{\hsize}{!}{\includegraphics[clip, trim=0 287 0 0 ]{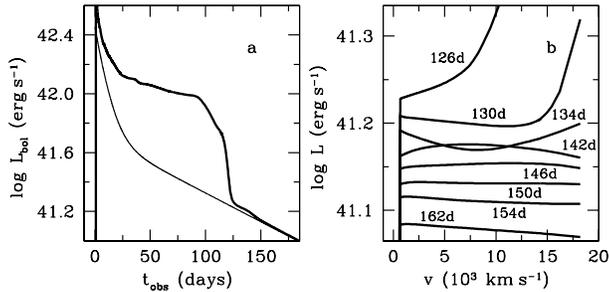}}
   \caption{%
   The plateau tail of the bolometric light curve in the optimal model.
   Panel~(\textbf{a}): calculated bolometric light curve (\emph{thick solid
      line\/}) is compared with the gamma-ray luminosity
      (\emph{thin solid line\/}).
   Panel~(\textbf{b}): evolution of the interior luminosity as a function of
      velocity.
   Profiles are plotted at $t=126$, 130, 134, 142, 146, 150, 154, and 162 days.
   }
   \label{fig:scndpl}
\end{figure}
By the end of the previous phase, the bolometric luminosity does not fall
   directly to the gamma-ray luminosity (Fig.~\ref{fig:scndpl}a),
   and the total radiation energy is not exhausted completely.
Nearly at the same time the photosphere disappears, and the SN envelope
   becomes optically thin for radiation.
For example, the total optical depth of the ejecta is 2.96 at the moment $t_5$,
   while the optical depth above the $^{56}$Ni bubble shell is only 0.78.
Because during the phase of the radiation energy exhaustion the photon gas in
   the outer layers cools more intensively than in the inner part of the envelope
   ($t=126$ and 130 days in Fig.~\ref{fig:scndpl}b), the radiation flow is generated
   in the warm inner layers ($t=130$ and 134 days in Fig.~\ref{fig:scndpl}b),
   then it propagates throughout the transparent layers, and results in some
   luminosity excess in the light curve just after the steep decline in the
   luminosity compared to the radioactive tail (Fig.~\ref{fig:scndpl}a).
We call this behavior of the light curve after the main plateau at the beginning
   of the radioactive tail ``a plateau tail''.

The radiation flow generated near the center of the envelope exists from
   $t \sim 130$ days to $t \sim 150$ days and then disappears
   (Fig.~\ref{fig:scndpl}b).
The characteristic duration of the plateau tail in the optimal model is 26
   days (Fig.~\ref{fig:scndpl}a).
Note that the energy excess radiated during the plateau tail over the luminosity
   of radioactive decays is $2.5 \times 10^{46}$ erg and $11\%$ of the total
   radiation energy within the ejecta at the moment $t_5$.

\subsection{Radioactive tail of the light curve}
\label{sec:dmd-rdtail}
After the phase of the plateau tail, the bolometric luminosity of the optimal
   model decreases to the gamma-ray luminosity due to the decay of the
   radioactive $^{56}$Co (Fig.~\ref{fig:scndpl}a) and is solely powered by
   the radioactive energy source.
This is how the stage of the radioactive tail starts, and the calculated light
   curve is in good agreement with the observations of SN 1999em after day 150
   (Fig.~\ref{fig:lmbol}).
Note that the envelope remains optically thick for the gamma rays during a few
   hundred days (the total optical depth is 30.6 at $t=150$ days), and
   they deposit the bulk of their energy locally, at least at the beginning of
   the radioactive tail.

\section{General properties of the optimal model}
\label{sec:genpro}
%
\begin{table}
\caption[]{Hydrodynamic models based on the optimal model D11.}
\label{tab:modsD11}
\centering
\begin{tabular}{c  l}
\hline\hline
\noalign{\smallskip}
 Model & Remarks \\
\noalign{\smallskip}
\hline
\noalign{\smallskip}
Dcc & no dense central core (Fig.~\ref{fig:denstr}a) \\
Ldn & less dense outer layers (Fig.~\ref{fig:denstr}c) \\
Mdn & more dense outer layers (Fig.~\ref{fig:denstr}c) \\
Hol & helium-rich composition in the outer layers: \\
    & $X=0.368$, $Y=0.615$, and $Z=0.017$ (Fig.~\ref{fig:sfchcm}a) \\
Zol & low metallicity in the outer layers: \\
    & $X=0.735$, $Y=0.259$, and $Z=0.006$ (Fig.~\ref{fig:sfchcm}c) \\
Sci & sharp metals/He/H composition interface (Fig.~\ref{fig:crchcm}a) \\
Hec & helium core of about 8 $M_{\sun}$ (Fig.~\ref{fig:crchcm}c) \\
Nhf & a half of $^{56}$Ni amount: $M_{\mathrm{Ni}}=0.018 M_{\sun}$ \\
Nno & no radioactive $^{56}$Ni \\
Nvm & $^{56}$Ni mixed up to a velocity of 830 km\,s$^{-1}$ \\
Nvh & $^{56}$Ni mixed up to a velocity of 1095 km\,s$^{-1}$ \\
Lop & no contribution of the expansion opacity \\
Ldk & without the limb-darkening effect\\
\noalign{\smallskip}
\hline
\end{tabular}
\end{table}
It is well known that the photometric characteristics of the SN~IIP outburst are
   determined mainly by the basic parameters: the pre-SN radius $R_0$,
   the ejecta mass $M_{env}$, and the explosion energy $E$.
Moreover, to get better agreement with the observations of SN 1999em, we have
   to vary the density distribution in the pre-SN model, its chemical
   composition, the $^{56}$Ni mass, and its mixing.
A dependence of the bolometric light curve on the expansion opacity and the
   effect of limb darkening is of particular interest.
These general properties of the optimal model are studied by means of the models
   listed in Table~\ref{tab:modsD11}.

\subsection{Presupernova structure}
\label{sec:grg-presn}
%
\begin{figure}[t]
   \resizebox{\hsize}{!}{\includegraphics{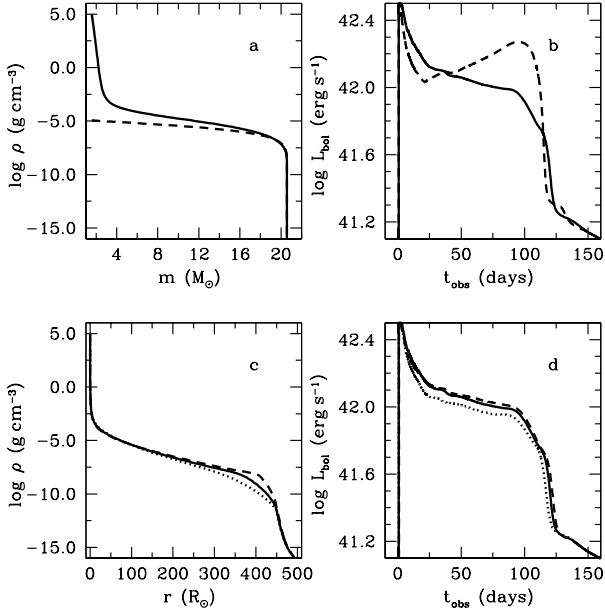}}
   \caption{%
   Dependence on the pre-SN structure.
   Top: (\textbf{a}) density distribution with respect to interior mass
      for the pre-SN model D11 (\emph{solid line\/}) and the
      similar model Dcc but without a dense central core (\emph{dashed line\/}),
      and (\textbf{b}) the corresponding bolometric light curves.
   Bottom: (\textbf{c}) density distribution with respect to radius
      for the pre-SN model D11 (\emph{solid line\/}) and the
      similar models Ldn and Mdn, but with less (\emph{dotted line\/}) and more
      (\emph{dashed line\/}) dense outer layers, respectively, and
      (\textbf{d}) the corresponding bolometric light curves.
   }
   \label{fig:denstr}
\end{figure}
At first, we study the dependence of the bolometric luminosity on the inner layers'
   density in the pre-SN model and consider the extreme case of an auxiliary
   model similar to the optimal model but without a dense central core
   (Fig.~\ref{fig:denstr}a).
In such a model the inner layers do less work moving outwards in the gravitational
   field than do those in the optimal model with a dense central core and hence
   acquire higher velocities than in the optimal model.
In contrast, the outer layers of the optimal model expand faster than those of
   the auxiliary model.
As a consequence, the photospheric radius of model D11 is greater than in
   the auxiliary model up to about day 35 and is smaller later on.
At the comparable effective temperatures of the models, this results in completely
   different behavior by the bolometric light curves (Fig.~\ref{fig:denstr}b).
This implies that there is a dense central core inside the real pre-SN of SN 1999em.

In terms of the CRW properties the growth in the bolometric luminosity during
   the CRW phase in the auxiliary model (Fig.~\ref{fig:denstr}b) is controlled
   by effective index $n$ at the photosphere position that is greater by
   $\sim 3.5-0.5$ than in the optimal model.
A greater effective index corresponds to a greater increase in the total flux
   of mass through the surface of the CRW front with time and, as a consequence,
   to faster growth in the bolometric luminosity with time (Grassberg \&
   Nadyozhin \cite{gn76}).

The influence of the density distribution in the outer layers of the pre-SN on
   the bolometric luminosity is illustrated by model D11 and the same models
   but with both less dense and denser outer layers (Figs.~\ref{fig:denstr}c
   and \ref{fig:denstr}d).
A roughly similar behavior of the effective index $n$ at the photosphere position
   in these models reflects the approximately equal slopes of the corresponding
   bolometric light curves during the CRW phase.
A transition from the model with the less dense outer layers to that of the
   denser layers is characterized by a reduction in the cooling by the
   adiabatic expansion and, consequently, by an increase in the bolometric
   luminosity in the whole SN outburst except for phases of the shock
   breakout and the radioactive tail.
As a result, for these models the energies radiated during the first 180 days
   are $1.251 \times 10^{49}$ erg, $1.355 \times 10^{49}$ erg, and
   $1.430 \times 10^{49}$ erg, respectively.
Thus, the bolometric light curve is fairly sensitive to the initial structure of
   the outer layers.

\subsection{Chemical composition}
\label{sec:grg-chcom}
%
\begin{figure}[t]
   \resizebox{\hsize}{!}{\includegraphics{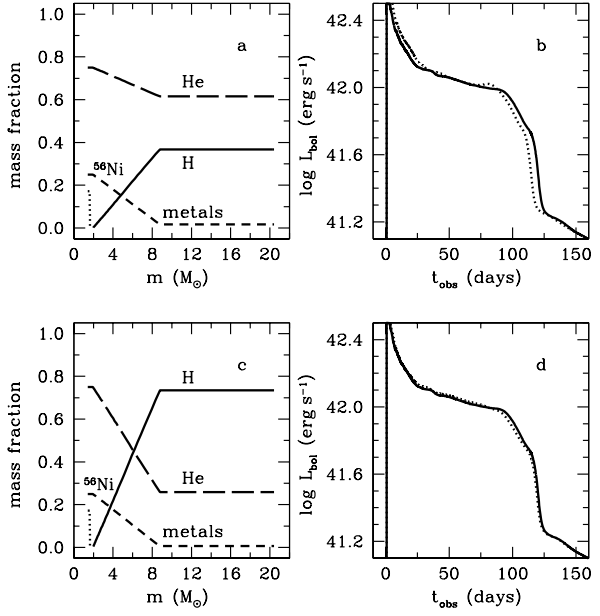}}
   \caption{%
   Dependence on the chemical composition in the outer layers of the ejecta
      beyond the helium core.
   Top: (\textbf{a}) the chemical composition of model Hol that differs
      from model D11 in the mass fraction of hydrogen (\emph{solid line\/})
      and helium (\emph{long dashed line\/}) (Table~\ref{tab:modsD11}),
      the distribution of heavy elements (\emph{short dashed line\/}) and
      radioactive $^{56}$Ni (\emph{dotted line\/}) being the same, and
      (\textbf{b}) the corresponding bolometric light curve (\emph{dotted line\/})
      and that of model D11 (\emph{solid line\/}).
   Bottom: (\textbf{c}) the chemical composition of model Zol that differs
      from model D11 by the mass fraction of helium and heavy elements
      (Table~\ref{tab:modsD11}), and (\textbf{d}) the corresponding bolometric
      light curve (\emph{dotted line\/}) and that of model D11 (\emph{solid line\/}).
   }
   \label{fig:sfchcm}
\end{figure}
In addition to the pre-SN structure, the chemical composition of the ejecta
   also affects the SN luminosity.
Let us consider a dependence on the chemical composition in the outer layers of
   the ejecta beyond the helium core.
An enhancement of helium abundance at the expense of hydrogen abundance in
   model Hol (Fig.~\ref{fig:sfchcm}a and Table~\ref{tab:modsD11}) results in
   the higher luminosity during the adiabatic cooling phase than that of
   model D11, nearly the same luminosity in the CRW phase, a small bump
   at the end of the plateau and then the shortened duration of the plateau
   (Fig.~\ref{fig:sfchcm}b).
This behavior of the luminosity is due to the lower opacity in the outer
   layers of model Hol.
On the other hand, lower metallicity for model Zol (Fig.~\ref{fig:sfchcm}c
   and Table~\ref{tab:modsD11}) than for model D11 (Table~\ref{tab:hydmods})
   increases the bolometric luminosity slightly during the CRW phase and then
   reduces it somewhat at the phase of radiative diffusion cooling
   (Fig.~\ref{fig:sfchcm}d).
This is explained by a smaller contribution of heavy elements to opacity in
   the outer layers of the ejecta in model Zol.
Thus, the bolometric light curve depends weakly on the chemical composition
   in the outer layers of the ejecta beyond the helium core, while hydrogen
   is abundant there and controls opacity of matter.

\begin{figure}[t]
   \resizebox{\hsize}{!}{\includegraphics{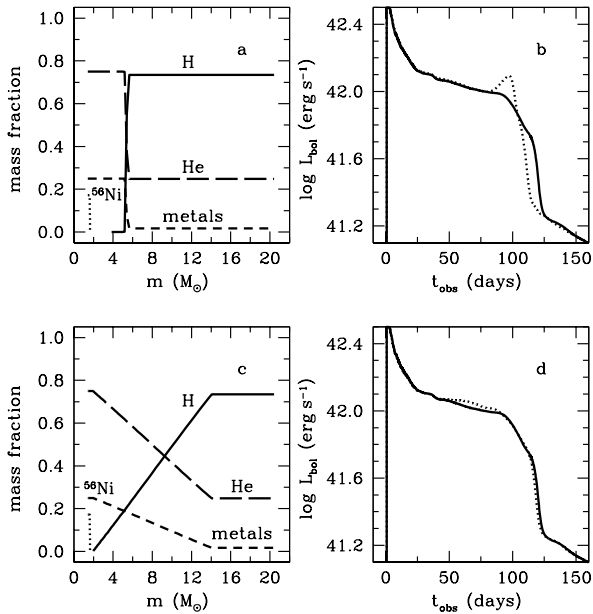}}
   \caption{%
   Dependence on the chemical composition in the inner layers of the ejecta.
   Top: (\textbf{a}) the chemical composition of model Sci, which differs
      from model D11 by the sharp metals/He/H composition interface, and
      (\textbf{b}) the corresponding bolometric light curve
      (\emph{dotted line\/}) and that of model D11 (\emph{solid line\/}).
   Bottom: (\textbf{c}) the chemical composition of model Hec, which differs
      from model D11 by the larger helium core, and
      (\textbf{d}) the corresponding bolometric light curve
      (\emph{dotted line\/}) and that of model D11 (\emph{solid line\/}).
   }
   \label{fig:crchcm}
\end{figure}
A good fit of the calculated bolometric light curve for the optimal model to
   what is observed (Fig.~\ref{fig:lmbol}) is obtained with mutual mixing of
   the hydrogen-rich and helium-rich matter in the inner layers of the ejecta
   (Fig.~\ref{fig:chcom}).
It is appropriate to ask here in what way the sharp boundary between
   the hydrogen-rich and helium-rich layers at the edge of the helium core ---
   a characteristic of the evolutionary models of pre-SNe --- changes the
   light curve.
In the auxiliary model Sci without the mutual mixing at the edge of the helium core
   (Table~\ref{tab:modsD11}) this boundary is located at the interior mass
   of 5.4 $M_{\sun}$ (Fig.~\ref{fig:crchcm}a).
It is known that a transition from the hydrogen-rich to helium-rich layers,
   i.e. from the low to high ionization potential matter, causes an increase in
   the effective temperature and, as a consequence, a growth in the bolometric
   luminosity (Grassberg \& Nadyozhin \cite{gn76}; Utrobin \cite{utr89}).
In fact, when the CRW front reaches the sharp boundary between the hydrogen-rich
   and helium-rich layers, crosses it, and then enters the helium core,
   a distinct bump in the light curve at the end of the plateau appears
   (Fig.~\ref{fig:crchcm}b).
Evidently, this local maximum of the luminosity is inconsistent with the observations
   of SN 1999em and other SNe IIP.

It is well known that at the time of an SN event the luminosity of
   a massive progenitor is determined mainly by the mass of the helium core,
   since a contribution of the hydrogen burning shell to the total energy
   generation is negligible.
Hence the mass of the helium core is a critical quantity in evaluating
   the evolutionary state of a massive star.
In this context it is intriguing that the increase in the helium core mass from
   5.6 $M_{\sun}$ in model D11 to 8.1 $M_{\sun}$ in model Hec
   (Fig.~\ref{fig:crchcm}c), under a deep hydrogen mixing downward to $\approx 700$
   km\,s$^{-1}$, leads to some decrease in the opacity of matter in the mixing
   region at the helium/hydrogen composition interface and, as a consequence,
   to the increase in the bolometric luminosity at the plateau (Fig.~\ref{fig:crchcm}d).
The resulting bolometric light curve is still consistent with the observations of
   SN 1999em.
Thus, the mass of the helium core between 5.6 $M_{\sun}$ and 8.1 $M_{\sun}$ is
   acceptable for the optimal model.

\subsection{Mass of {\boldmath $^{56}$}Ni and its mixing}
\label{sec:grg-nicl}
%
\begin{figure}[t]
   \resizebox{\hsize}{!}{\includegraphics[clip, trim=0 287 0 0 ]{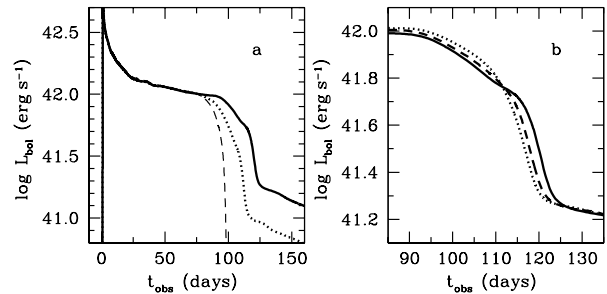}}
   \caption{%
   Influence of the $^{56}$Ni mass and its mixing.
   Panel~(\textbf{a}): bolometric light curves for model D11 (\emph{solid line\/})
      and the similar models Nhf and Nno but calculated with a half of the $^{56}$Ni
      amount (\emph{dotted line\/}) and without $^{56}$Ni (\emph{dashed line\/}),
      respectively.
   Panel~(\textbf{b}): bolometric light curves at the end of the plateau for
      model D11 (\emph{solid line\/}), model Nvm (\emph{dashed line\/}), and
      model Nvh (\emph{dotted line\/}) in which $^{56}$Ni is mixed up to a velocity
      of 660, 830, and 1095 km\,s$^{-1}$, respectively.
   }
   \label{fig:nimsmx}
\end{figure}
After the CRW stage, when the radiative diffusion takes place, the radioactive
   decay of the $^{56}$Ni and $^{56}$Co nuclides begins to dominate in powering
   the luminosity.
This fact is clearly demonstrated by model Nno without $^{56}$Ni in the envelope
   (Fig.~\ref{fig:nimsmx}a and Table~\ref{tab:modsD11}).
A lack of $^{56}$Ni shortens the duration of the bolometric light curve to about
   100 days.
It is evident that at later times the bolometric light curve depends on both the
   total mass of $^{56}$Ni and its distribution over the ejecta.
Of course, the total mass of $^{56}$Ni is measured by the radioactive tail of
   the observed light curve and is 0.036 $M_{\sun}$ in the optimal model for
   the Cepheid distance of 11.7 Mpc.
A reduction in the total mass of $^{56}$Ni from 0.036 $M_{\sun}$ (model D11 in
   Table~\ref{tab:hydmods}) to 0.018 $M_{\sun}$ (model Nhf in Table~\ref{tab:modsD11})
   results mainly in a shorter duration of the light curve (Fig.~\ref{fig:nimsmx}a).
Thus, the duration of the bolometric light curve has a minimum value of about
   100 days and is a function of the $^{56}$Ni amount.

To investigate the influence of the $^{56}$Ni mixing on the light curve, hydrodynamic
   models Nvm and Nvh were calculated with the $^{56}$Ni distribution different from
   that of model D11, but containing the same total mass of $^{56}$Ni
   (Table~\ref{tab:modsD11}).
In these models the radioactive nickel is spread over a larger mass range of
   the ejected envelope or, equivalently, over a wider velocity range
   than in model D11.
Hence, more favorable conditions for the radiative diffusion and an additional
   heating due to the radioactive decays should appear and therefore should
   increase the luminosity at the phase of radiative diffusion cooling that is
   clearly seen in Fig.~\ref{fig:nimsmx}b.
In addition, this increase in the luminosity causes its subsequently earlier
   fall to the radioactive tail.
As a result, with increasing the $^{56}$Ni mixing in velocity space the linear
   luminosity shoulder at the phase of radiative diffusion cooling transforms
   into the convex light curve at the end of the plateau.
Thus, a linear shoulder of the luminosity in logarithmic scale at the phase of
   radiative diffusion cooling implies a weak $^{56}$Ni mixing.
Because a discrepancy between the light curve of model D11 (which is in agreement
   with the observations of SN 1999em) and those of models Nvm and Nvh is rather
   significant, one can conclude that the bulk of the radioactive $^{56}$Ni
   should be confined to the innermost layers of the ejected envelope expanding
   with velocities less than 660 km\,s$^{-1}$.

\subsection{Plateau tail}
\label{sec:grg-secnd}
%
\begin{figure}[t]
   \resizebox{\hsize}{!}{\includegraphics[clip, trim=0 287 0 0 ]{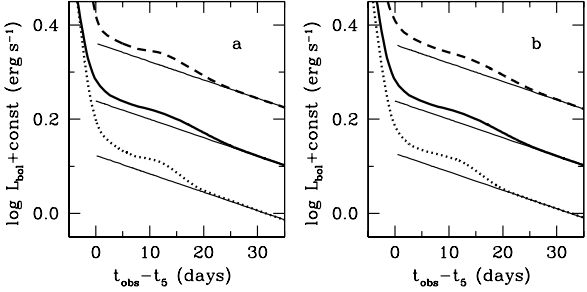}}
   \caption{%
   Dependence of the plateau tail of the bolometric light curve on (\textbf{a})
      the $^{56}$Ni amount (model Nhf in Table~\ref{tab:modsD11}, \emph{dotted
      line\/}) and the contribution of the expansion opacity (model Lop in
      Table~\ref{tab:modsD11}, \emph{dashed line\/}), and on (\textbf{b})
      the ejecta mass (model Mps in Table~\ref{tab:auxmods}, \emph{dotted line\/})
      and the explosion energy (model Eps in Table~\ref{tab:auxmods},
      \emph{dashed line\/}).
   \emph{Thick solid line} is the bolometric light curve of model D11
      and \emph{thin solid line} is the luminosity of radioactive decays.
   }
   \label{fig:gpsdpl}
\end{figure}
The general properties of the plateau tail on the radioactive tail are
   illustrated by the light curves shown in Fig.~\ref{fig:gpsdpl} and the
   relevant parameters of the models listed in Table~\ref{tab:bp2plt}
   where $t_5$ is the time of the onset of the plateau tail as defined in
   Fig.~\ref{fig:snphs}, $\Delta t^{2p}$ the characteristic duration of
   the plateau tail, $\Delta t_R = R_{ext}/c$ is the characteristic time
   of radiation flow for the external radius of the expelled envelope $R_{ext}$
   in the middle of the plateau tail, $\tau_{tot}$ the total optical
   depth of the ejecta at the moment $t_5$, $\Delta E_r^{2p}$ the energy
   excess radiated during the plateau tail over the luminosity of radioactive
   decays, and $\Delta E_r^{2p}/E_r^{tot}$ is a ratio of this energy to
   the total radiation energy within the ejecta at the moment $t_5$.

\begin{table}
\caption[]{Basic parameters at the phase of the plateau tail.}
\label{tab:bp2plt}
\centering
\begin{tabular}{c @{ } c @{ } c @{ } c @{ } c @{ } c  @{ } c  @{ } c}
\hline\hline
\noalign{\smallskip}
 Model & $t_5$ & $\Delta t^{2p}$ & $\Delta t_R$ & $\Delta t^{2p}/\Delta t_R$ &
         $\tau_{tot}$ & $\Delta E_r^{2p}$ & $\Delta E_r^{2p}/E_r^{tot}$ \\
       & (days) & (days) & (days) & & & ($10^{46}$ erg) & \\
\noalign{\smallskip}
\hline
\noalign{\smallskip}
D11 & 124 & 26 & 8.76 & 2.97 & 2.96 & 2.5 & 0.110 \\
Lop & 110 & 25 & 7.83 & 3.19 & 1.71 & 3.8 & 0.136 \\
Nhf & 115 & 22 & 7.94 & 2.77 & 2.99 & 1.5 & 0.100 \\
Mps & 127 & 22 & 7.89 & 2.79 & 2.64 & 2.3 & 0.089 \\
Eps & 119 & 28 & 9.25 & 3.03 & 3.01 & 3.1 & 0.107 \\
\noalign{\smallskip}
\hline
\end{tabular}
\end{table}
It is clear that there is no correlation between the characteristic duration
   of the plateau tail and the total optical depth of the ejecta
   (Table~\ref{tab:bp2plt}), and the duration depends very weakly on the
   opacity of matter (models D11 and Lop in Table~\ref{tab:bp2plt}).
These regularities reflect the fact that the radiation responsible for the
   plateau tail flows throughout an optically thin medium.
On the other hand, the characteristic duration is proportional to the
   characteristic time of radiation flow independent of the parameters of
   the listed models (Table~\ref{tab:bp2plt}), and their ratio is 2.95 on
   average.
It is worth noting that the energy excess of the plateau tail is roughly
   $10 \%$ of the total radiation energy at the moment $t_5$ for all
   but model Lop (Table~\ref{tab:bp2plt}).

The plateau tail of the bolometric light curve demonstrates the following
   important properties.
A decrease in the $^{56}$Ni mass (models D11 and Nhf) shortens the
   characteristic duration of the plateau tail and reduces its energy
   excess (Fig.~\ref{fig:gpsdpl}a and Table~\ref{tab:bp2plt}).
This influence of the $^{56}$Ni amount is related to its large role in
   powering the light curve at the end of the main plateau after day 100
   (Fig.~\ref{fig:nimsmx}a).
The higher the average expansion velocity of the ejecta (models Mps, D11,
   and Eps), the longer the duration of the plateau tail and the higher
   the energy excess (Fig.~\ref{fig:gpsdpl}b and Table~\ref{tab:bp2plt}).
An increase in the duration of the plateau tail with the average expansion
   velocity results from its proportionality to the characteristic time of
   radiation flow.
The higher average expansion velocity causes the onset of the plateau tail
   at an earlier time, when the total radiation energy is higher and, as
   a consequence, the energy excess, radiated over the luminosity of
   radioactive decays during the plateau tail, becomes greater.

\subsection{Expansion opacity}
\label{sec:grg-lop}
%
\begin{figure}[t]
   \resizebox{\hsize}{!}{\includegraphics[clip, trim=0 0 0 184]{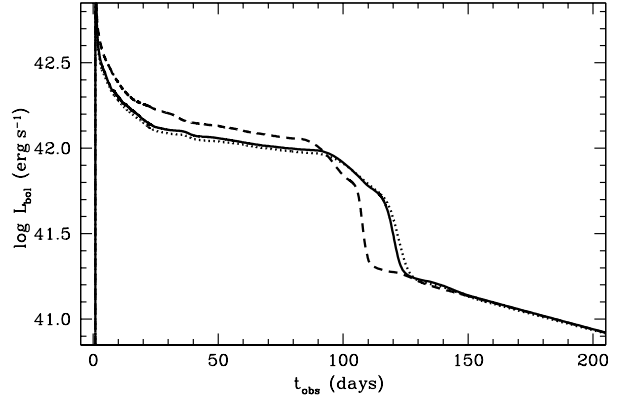}}
   \caption{%
   The bolometric light curves for model D11 (\emph{solid line\/}) and
      the similar models Lop and Ldk, but calculated without the expansion
      opacity (\emph{dashed line\/}), and for the isotropic emergent radiation,
      without the limb darkening (\emph{dotted line\/}).
   }
   \label{fig:lopldk}
\end{figure}
A contribution of numerous metal lines to opacity plays a fundamental role
   in reproducing the observed light curve of SN 1999em, a normal SN IIP,
   as shown by the optimal model D11 and model Lop calculated by neglecting
   the expansion opacity (Fig.~\ref{fig:lopldk}).
The decrease in opacity due to disregarding the line contribution in model Lop
   speeds up the radiation diffusion and causes the bolometric luminosity
   to increase significantly compared to that of model D11 during nearly
   the whole outburst and, as a consequence, the characteristic duration
   of the plateau to shorten by about 10 days.
It is equivalent in action to an increase in the explosion energy that will
   be demonstrated in Sect.~\ref{sec:phyobs} (Fig.~\ref{fig:baspar}c).
This fact leads to the firm conclusion that neglecting the expansion opacity
    underestimates the explosion energy and introduces an error of nearly
    $20\%$ into its value.

It is worth noting that the slope of the light curve shoulder at the phase of
   radiative diffusion cooling in model D11 is flatter than in model Lop
   without the expansion opacity (Fig.~\ref{fig:lopldk}).
This behavior of the light curve is consistent with the dependence of the
   characteristic diffusion time on the optical depth of the diffusion region
   (\ref{eq:lboldif}) which, in turn, is proportional to the opacity of matter.

\begin{figure}[t]
   \resizebox{\hsize}{!}{\includegraphics{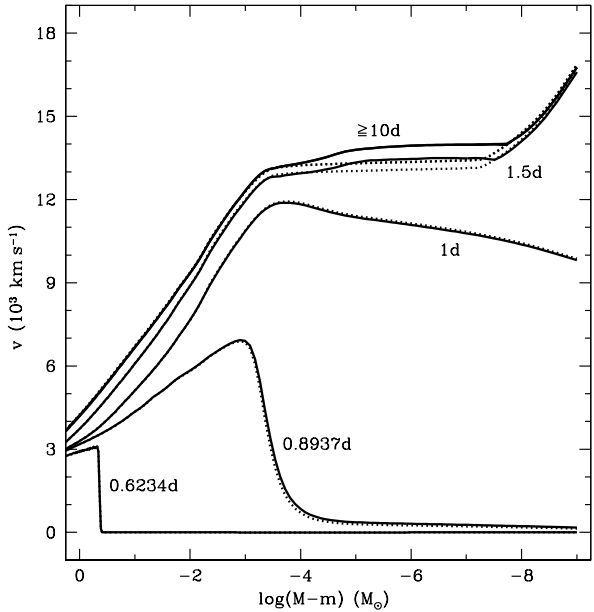}}
   \caption{%
   Evolution of the velocity distribution in mass for models D11
      (\emph{solid line\/}) and Lop (\emph{dotted line\/}):
      at $t=0.6234$ days when the shock approaches the pre-SN surface,
      at $t=0.8937$ days when the shock emerges on the surface,
      at $t=1$ days, at $t=1.5$ days, and subsequently, at later times
      $t \geq 10$ days when the envelope is already expanding homologously.
   The mass is measured from the envelope surface.
   }
   \label{fig:lopvel}
\end{figure}
The resonance scattering of radiation in numerous metal lines is also important
   in forming the gas flow in the outer layers of the expanding SN
   envelope (Fig.~\ref{fig:lopvel}).
After the shock emerges on the pre-SN surface at $t=0.8937$ days, an
   expansion of the envelope begins, and it becomes homologous during the
   next 10 days.
An additional acceleration induced by the expansion opacity occurs between $t \approx 1$
   days and $t \approx 1.5$ days, lasts up to $t \approx 7.5$ days, and takes place
   in the outer layers with a mass of about $10^{-4} M_{\sun}$.
A comparison of models D11 and Lop shows that these layers gain a velocity excess
   as large as 600 km\,s$^{-1}$, except for the outermost layers with a mass of
   about $10^{-8} M_{\sun}$.
Note that this additional acceleration is responsible for a feature in the gas
   flow as seen in Fig.~\ref{fig:freexp} where the velocities of different
   mass shells in the ejected envelope as a function of time are overplotted
   on the photospheric velocity.

\subsection{Effect of limb darkening}
\label{sec:grg-lmbdkg}
During the first 116 days, when a well-defined photosphere exists, particularly
   at the CRW phase, the emitted radiation is nearly isotropic, and the
   bolometric light curves of models D11 and Ldk, calculated without the
   effect of the limb darkening, almost coincide (Fig.~\ref{fig:lopldk}).
As the envelope expands and becomes optically thin, the continuum formation
   region gradually becomes more extended, and the degree of anisotropy of
   the emergent radiation increases.
The increase in the radiation anisotropy with time clearly results in the
   difference between the light curves of models D11 and Ldk during the phase
   of the radiation energy exhaustion: the bolometric luminosity calculated
   by taking the limb-darkening law into account is lower than for
   isotropic radiation (Fig.~\ref{fig:lopldk}).

Utrobin (\cite{utr04}) shows that the difference between the bolometric
   luminosity calculated by taking into account the emergent anisotropic
   radiation and the luminosity for the isotropic radiation depends on both
   the limb-darkening law and the retardation effect.
Interestingly, this difference increases with growing the degree of the anisotropy
   of the emergent radiation, while its sign is determined solely by the time
   derivative of the bolometric luminosity.
The latter property explains the behavior of the bolometric luminosity for models
   D11 and Ldk discussed above.
It seems that in SN 1999em, a normal SN IIP, the effect of limb darkening is
   more of purely academic interest than of practical value.

\section{Physical and observed parameters}
\label{sec:phyobs}
%
\begin{figure}[t]
   \resizebox{\hsize}{!}{\includegraphics[clip, trim=0 0 0 184]{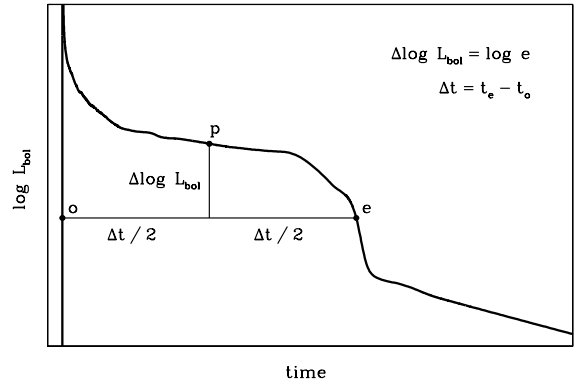}}
   \caption{%
   Schematic bolometric light curve of SN IIP.
   Lettered dots mark the onset of the luminosity rising (o), the middle of
      the plateau (p), and the end of the plateau (e).
   }
   \label{fig:bscshm}
\end{figure}
Generally, our main goal is to construct the adequate hydrodynamic model by fitting
   the photometric and spectroscopic observations of the object under study.
Sometimes it is tempting to evaluate the relevant physical parameters for other
   SN with close observed properties.
In this case, it is reasonable to use the results already obtained for the similar
   object.
For this purpose, we have to find the relationships between the basic physical and
   observed parameters.
Evidently, the important physical parameters are the initial radius $R_0$, the
   ejecta mass $M_{env}$, and the explosion energy $E$.
The large impact of $^{56}$Ni on the light curve near the end of the plateau,
   which is shown and discussed in Sect.~\ref{sec:grg-nicl}, makes us take the
   total $^{56}$Ni mass into account.
Because of the temporal character of the energy deposition from radioactive decays,
   it is impossible to treat this energy input as an addition to the explosion
   energy, so we have to introduce the total $^{56}$Ni mass $M_{\mathrm{Ni}}$
   as the fourth basic physical parameter.

\begin{figure}[t]
   \resizebox{\hsize}{!}{\includegraphics{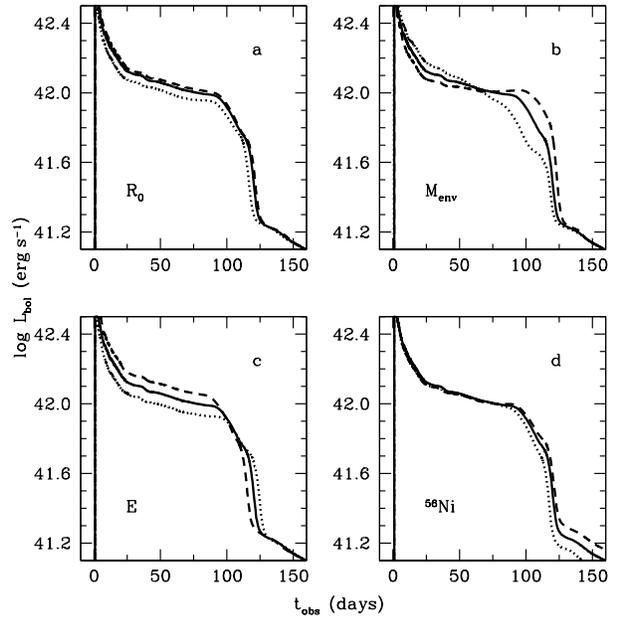}}
   \caption{%
   Dependence of the bolometric light curve of the optimal model D11
      (\emph{solid line\/}) on the basic parameters:
      (\textbf{a}) the initial radius, model Rms (\emph{dotted line\/})
      and model Rps (\emph{dashed line\/});
      (\textbf{b}) the ejecta mass, model Mms (\emph{dotted line\/})
      and model Mps (\emph{dashed line\/});
      (\textbf{c}) the explosion energy, model Ems (\emph{dotted line\/})
      and model Eps (\emph{dashed line\/});
      (\textbf{d}) the total $^{56}$Ni mass, model Nms
      (\emph{dotted line\/}) and model Nps (\emph{dashed line\/}).
   }
   \label{fig:baspar}
\end{figure}

To describe the observed properties of SN IIP, Litvinova \& Nadyozhin (\cite{ln83},
   \cite{ln85}) used the three parameters: the plateau duration, the absolute V
   magnitude, and the photospheric velocity in the middle of the plateau.
However, in the course of time it became evident that this simple description
   needed to be corrected for the influence of $^{56}$Ni on the light curve via
   an additional parameter (Nadyozhin \cite{n03}).
Unfortunately, the bounds of the plateau are unclear because they are no
   specific points on the light curve.
Moreover, the interval of a plateau duration misses the bolometric luminosity jump
   at the shock breakout and the subsequent adiabatic cooling phase together with
   the relevant dependence on the initial radius, the ejecta mass, and the explosion
   energy (Fig.~\ref{fig:baspar}).
To overcome this problem, instead of the plateau duration we define a characteristic
   duration $\Delta t$ of the light curve as a whole, measuring it from the
   bolometric luminosity jump at the shock breakout to the end of the plateau for
   an $e$-fold reduction of the luminosity in the middle of plateau
   (Fig.~\ref{fig:bscshm}).
Of course, in this way the characteristic duration is easy to determine for
   a theoretical light curve, as shown in a schematic bolometric light curve of
   SN IIP, but very difficult for an observed one.
Fortunately, this situation is not hopeless.
First, it is possible to apply the theoretical light curve as a template for
   measuring the specific moments $t_o$ and $t_e$ (Fig.~\ref{fig:bscshm}) and,
   consequently, for calculating the characteristic duration.
Then, as the observational data become more and more extensive, the relevant
   templates may be constructed from the observed light curves of the well-studied
   SNe IIP.
The second modification we made is to use the bolometric luminosity
   $L_{bol}^{p}$ in the middle of the plateau because it is more adequate
   than the absolute V magnitude for estimating the explosion energy.
The photospheric velocity $v_{ph}^{p}$ in the middle of the plateau is the third
   observed parameter in accordance with Litvinova \& Nadyozhin (\cite{ln83},
   \cite{ln85}).

\begin{table}
\caption[]{Physical parameters of auxiliary hydrodynamic models.}
\label{tab:auxmods}
\centering
\begin{tabular}{c  c  c @{ } c @{ }  c  c  c}
\hline\hline
\noalign{\smallskip}
 Model & $R_0$ & $M_{env}$ & $E$ & $M_{\mathrm{Ni}}$ & $X$ & $Z$ \\
       & ($R_{\sun}$) & ($M_{\sun}$) & ($10^{51}$ erg) & $(10^{-2} M_{\sun})$ &
       & \\
\noalign{\smallskip}
\hline
\noalign{\smallskip}
D11 & 500 & 19 & 1.30 & 3.60 & 0.735 & 0.017 \\
Rms & 425 & 19 & 1.30 & 3.60 & 0.735 & 0.017 \\
Rps & 575 & 19 & 1.30 & 3.60 & 0.735 & 0.017 \\
Mms & 500 & 16 & 1.30 & 3.60 & 0.735 & 0.017 \\
Mps & 500 & 22 & 1.30 & 3.60 & 0.735 & 0.017 \\
Ems & 500 & 19 & 1.10 & 3.60 & 0.735 & 0.017 \\
Eps & 500 & 19 & 1.50 & 3.60 & 0.735 & 0.017 \\
Nms & 500 & 19 & 1.30 & 3.06 & 0.735 & 0.017 \\
Nps & 500 & 19 & 1.30 & 4.14 & 0.735 & 0.017 \\
\noalign{\smallskip}
\hline
\end{tabular}
\end{table}
To establish the relationships among the basic physical and observed parameters accepted
   above, we carried out a parameter study of hydrodynamic models varying the
   physical parameters and measuring the observed ones.
Evidently, a four-parameter approximation is a crude description of the hydrodynamic
   model properties.
The sensitivity of the bolometric light curve to the pre-SN structure and
   its chemical composition makes this approximation valid only for the vicinity
   of the optimal model D11, not for a whole region of the basic parameters.
We restricted ourselves to a local parameter study in the vicinity of the optimal
   model and explored a $15\%$ limited range of the basic physical parameters
   (Table~\ref{tab:auxmods}).
The resulting bolometric light curves of auxiliary hydrodynamic models are plotted
   in Fig.~\ref{fig:baspar} and compared to the optimal model.
The basic observed parameters of these models are evaluated according to the
   definition, given in Fig.~\ref{fig:bscshm}, and listed in Table~\ref{tab:pobpar}.

The influence of the basic physical parameters is illustrated by the bolometric
   light curves of the auxiliary models (Fig.~\ref{fig:baspar}).
Increasing the initial radius causes an increase in the characteristic time of
   the envelope expansion and, as a consequence, an increase in the width of
   the narrow intense peak on the bolometric light curve produced by the heating
   of the outer layers at the shock breakout phase.
In addition, an increase in the initial radius leads to a reduction in the
   cooling by the adiabatic expansion and, accordingly, to an increase in the
   bolometric luminosity during the whole SN outburst except for the phase
   of the radioactive tail (Fig.~\ref{fig:baspar}a).
It is interesting that this behavior by the bolometric light curve with initial
   radius is similar to its dependence on the density structure in the outer
   layers (Fig.~\ref{fig:denstr}c), which is discussed in Sect.~\ref{sec:grg-presn}.
The latter dependence may be considered as a dependence on some effective radius
   of the pre-SN.
In other words, an increase in the density in the outer layers of the pre-SN
   mimics an increase in the initial radius.
Both decreasing the ejecta mass and increasing the explosion energy enlarge the
   average expansion velocity of the envelope.
The increase in the average velocity of the envelope makes the shock
   wave propagating through the pre-SN matter stronger and, as a result, it heats
   the matter to a higher temperature, which increases the luminosity in the
   narrow peak of the light curve, at the subsequent phase of adiabatic cooling
   and at the beginning of the CRW phase (Figs.~\ref{fig:baspar}b and
   \ref{fig:baspar}c).
By the end of the CRW phase, decreasing the ejecta mass reduces the bolometric
   luminosity and shortens the characteristic duration of the light curve
   (Fig.~\ref{fig:baspar}b).
In contrast, increasing the explosion energy results in the luminosity growing
   and, accordingly, in shortening the characteristic duration as well
   (Fig.~\ref{fig:baspar}c).
Up to the phase of the radiative diffusion cooling, the radioactive $^{56}$Ni does
   not affect the light curve; but during this phase and later on, the energy
   deposition from radioactive decays powers the luminosity (Fig.~\ref{fig:baspar}d).
The higher the total $^{56}$Ni mass, the greater the energy deposition and the
   later the luminosity descending to the radioactive tail.
Note that a slope of the luminosity shoulder at the phase of the radiative
   diffusion cooling clearly depends on the initial radius (Fig.~\ref{fig:baspar}a)
   and on the explosion energy (Fig.~\ref{fig:baspar}c) in a qualitative agreement
   with the simple description of photon diffusion (\ref{eq:lboldif}).

\begin{table}
\caption[]{Observed parameters of auxiliary hydrodynamic models.}
\label{tab:pobpar}
\centering
\begin{tabular}{c c c c c c}
\hline\hline
\noalign{\smallskip}
 Model & $\Delta t$ & $\log L_{bol}^{p}$ & $v_{ph}^{p}$ & $q_{kin}$ & $q_{rad}$ \\
       & (days) & (erg\,s$^{-1}$) & (km\,s$^{-1}$) & & \\
\noalign{\smallskip}
\hline
\noalign{\smallskip}
D11 & 117.57 & 42.0350 & 2374.4 & 0.8541 & 0.8258 \\
Rms & 114.88 & 41.9927 & 2335.3 & 0.8088 & 0.8429 \\
Rps & 118.25 & 42.0521 & 2388.3 & 0.8354 & 0.7826 \\
Mms & 111.70 & 42.0542 & 2481.5 & 0.7767 & 0.7881 \\
Mps & 121.64 & 42.0170 & 2291.9 & 0.9141 & 0.8252 \\
Ems & 122.50 & 41.9629 & 2134.7 & 0.8134 & 0.8040 \\
Eps & 111.93 & 42.0979 & 2640.5 & 0.8890 & 0.8424 \\
Nms & 113.98 & 42.0372 & 2452.5 & 0.8943 & 0.8303 \\
Nps & 119.39 & 42.0264 & 2320.3 & 0.7968 & 0.8102 \\
\noalign{\smallskip}
\hline
\end{tabular}
\end{table}
To evaluate the three basic physical parameters (the initial radius, the ejecta mass,
   and the explosion energy) from the four values measured from observations
   (the characteristic duration of the light curve, the bolometric luminosity and
   the photospheric velocity in the middle of the plateau, and the total $^{56}$Ni
   mass), we use the following approximate relations:
\begin{eqnarray}
   \log R_0 & = & + 3.481 \log L_{bol}^{p} - 5.937 \log v_{ph}^{p}
                  - 1.999 \log \Delta t \nonumber \\
   &   & - 0.499 \log M_{\mathrm{Ni}} - 120.174 \; ,
\label{eq:logr}
\end{eqnarray}
\begin{eqnarray}
   \log M_{env} & = & - 2.942 \log L_{bol}^{p} + 7.606 \log v_{ph}^{p}
                  + 7.807 \log \Delta t \nonumber \\
   &   & - 0.042 \log M_{\mathrm{Ni}} + 83.045 \; ,
\label{eq:logm}
\end{eqnarray}
\begin{eqnarray}
   \log E & = & - 1.476 \log L_{bol}^{p} + 4.899 \log v_{ph}^{p}
                + 3.040 \log \Delta t \nonumber \\
   &   & + 0.312 \log M_{\mathrm{Ni}} + 90.765 \; ,
\label{eq:loge}
\end{eqnarray}
   where $R_0$, $M_{env}$, and $M_{\mathrm{Ni}}$ are in solar units, $E$ in
   units of erg, $L_{bol}^{p}$ in erg\,s$^{-1}$, $v_{ph}^{p}$ in km\,s$^{-1}$,
   and $\Delta t$ in days.
The above approximate formulae (\ref{eq:logr}--\ref{eq:loge}) are found by means
   of minimizing the errors between the values calculated with these relations
   and the corresponding physical parameters from Table~\ref{tab:auxmods}.
For this purpose we constructed a variational functional in a quadratic
   form depending on the functional values, i.e. coefficients in the approximate
   formulae (\ref{eq:logr}--\ref{eq:loge}), and then used the direct search
   method of Powell (\cite{p64}).
Note that the obtained approximate formulae give the estimates of the physical
   parameters for the auxiliary models, the accurate values of which are listed
   in Table~\ref{tab:auxmods}, with an error as great as $\approx 2.5\%$.

In addition to the observed parameters of the auxiliary models,
   the important energetic ratios
   $q_{kin} = 0.5 M_{env} {v_{ph}^{p}}^2 / E_{kin}$ and
   $q_{rad} = L_{bol}^{p} \Delta t / E_{rad}$, where $E_{kin}$ is the kinetic
   energy of the ejecta and $E_{rad}$ is the total radiation energy emitted during
   the first 180 days, are given in Table~\ref{tab:pobpar}.
The mean values of the ratios $q_{kin}$ and $q_{rad}$ are 0.842 and 0.816 with
   the standard deviations of $4.4\%$ and $2.3\%$, respectively.
A quite low value of the standard deviation in the energetic ratios shows that
   the kinetic energy of the envelope and the total radiation energy are
   approximated more or less accurately by the observed parameters: $L_{bol}^{p}$,
   $v_{ph}^{p}$, and $\Delta t$.

A functional dependence of the observed parameters on the physical ones is,
   in turn, derived directly from the approximate formulae
   (\ref{eq:logr}--\ref{eq:loge}) and can be written as
\begin{eqnarray}
   \log L_{bol}^{p} & = & + 0.488 \log R_0 - 0.267 \log M_{env}
                  + 1.006 \log E \nonumber \\
   &   & - 0.082 \log M_{\mathrm{Ni}} - 10.469 \; ,
\label{eq:logl}
\end{eqnarray}
\begin{eqnarray}
   \log v_{ph}^{p} & = & + 0.083 \log R_0 - 0.247 \log M_{env}
                  + 0.688 \log E \nonumber \\
   &   & - 0.184 \log M_{\mathrm{Ni}} - 31.951 \; ,
\label{eq:logv}
\end{eqnarray}
\begin{eqnarray}
   \log \Delta t & = & + 0.103 \log R_0 + 0.268 \log M_{env}
                  - 0.291 \log E \nonumber \\
   &   & + 0.154 \log M_{\mathrm{Ni}} + 16.548 \; .
\label{eq:logdt}
\end{eqnarray}
The estimates of the observed parameters for the auxiliary models obtained with
   the above relations (\ref{eq:logl}--\ref{eq:logdt}) differ from the accurate
   values given in Table~\ref{tab:pobpar} by an error less than $\approx 2.2\%$.

Finally, it should be emphasized that all of the above formulae approximate the
   physical parameters and the observable properties of hydrodynamic models
   but only in the vicinity of the optimal model D11.
To cover a wider range of the parameters, we have to construct a global
   approximation, which should be based, in our opinion, on a set of hydrodynamic
   models of \emph{many real} SNe IIP, but not on that of arbitrary hydrodynamic
   models.

\section{Comparison with SN 1987A}
\label{sec:sn87a}
%
\begin{table}[t]
\caption[]{Hydrodynamic models for SN 1999em and SN 1987A.}
\label{tab:hmsn99sn87}
\centering
\begin{tabular}{@{ } c @{ } c @{ } c @{ } c @{ } c @{ } c @{ } c @{ } c @{ }}
\hline\hline
\noalign{\smallskip}
 SN & $R_0$ & $M_{env}$ & $E$ & $M_{\mathrm{Ni}}$ & $Z$
       & $v_{\mathrm{Ni}}^{max}$ & $v_{\mathrm{H}}^{min}$ \\
       & $(R_{\sun})$ & $(M_{\sun})$ & ($10^{51}$ erg) & $(10^{-2} M_{\sun})$
       & & (km\,s$^{-1}$) & (km\,s$^{-1}$) \\
\noalign{\smallskip}
\hline
\noalign{\smallskip}
99em & 500 & 19 & 1.3 & 3.60 & 0.017 &  660 & 700 \\
87A &  35 & 18 & 1.5 & 7.65 & 0.006 & 3000 & 600 \\
\noalign{\smallskip}
\hline
\end{tabular}
\end{table}
Now we have got a confidence in reproducing the observed bolometric light
   curve and the H$\alpha$ line of SN 1999em, a normal SN IIP, we can compare this
   object with the well-studied SN 1987A, a peculiar SN IIP.
The important parameters of the optimal model D11 for SN 1999em and model M18
   for SN 1987A (Utrobin \cite{utr05}) are given in Table~\ref{tab:hmsn99sn87}.
It is remarkable that the ejecta mass $M_{env}$, the explosion energy $E$, and
   the minimum velocity of the hydrogen-rich envelope $v_{\mathrm{H}}^{min}$ are
   comparable, while the pre-SN radius $R_0$, the mass fraction of heavy
   elements in the outer layers of the pre-SN $Z$, the total mass of
   $^{56}$Ni $M_{\mathrm{Ni}}$, and its maximum velocity $v_{\mathrm{Ni}}^{max}$
   are quite different.

The mass of a star is a fundamental parameter that determines the properties
   of the star and the course of its evolution.
The pre-SN masses of SN 1999em and SN 1987A, estimated by the mass of
   the neutron star and the ejecta mass, are nearly 20.6 $M_{\sun}$ and
   19.6 $M_{\sun}$, respectively.
The masses of helium cores in the pre-SNe of SN 1999em and SN 1987A are
   5.6--8.1 $M_{\sun}$ and 6.0 $M_{\sun}$ (Woosley \cite{w88}), respectively.
These masses are close enough to suppose that, in the final stages of stellar
   evolution, nearly the same iron cores form within the pre-SNe.
This fact and roughly the same explosion energies of SN 1999em and SN 1987A
   imply a unique explosion mechanism for these core collapse SNe.

It is evident that the amount of radioactive $^{56}$Ni and its distribution
   in the SN envelope are a clear trace left by the explosion mechanism.
Both SN 1999em and SN 1987A differ in the total mass of radioactive
   $^{56}$Ni and in its distribution throughout the envelope
   (Table~\ref{tab:hmsn99sn87}).
Note that for these SNe the total $^{56}$Ni mass is correlated to the
   explosion energy.
This dependence is consistent with the empirical correlation between the total
   $^{56}$Ni mass and the explosion energy for SNe IIP found by Nadyozhin
   (\cite{n03}) and Hamuy (\cite{ham03}).
The optimal model for SN 1999em is characterized by a weak $^{56}$Ni mixing in
   velocity space up to $\approx 660$ km\,s$^{-1}$ and a deep hydrogen mixing
   downward to $\approx 700$ km\,s$^{-1}$ (Fig.~\ref{fig:chcom} and
   Table~\ref{tab:hmsn99sn87}).
In turn, the hydrodynamic model for SN 1987A is characterized by a moderate
   $^{56}$Ni mixing up to $\sim 3000$ km\,s$^{-1}$ and a deep hydrogen mixing
   downward to $\sim 600$ km\,s$^{-1}$ (Utrobin \cite{utr05}).

The structure of the pre-SN and the chemical composition of its outer
   layers, which are unaffected by explosive nucleosynthesis and which are
   the end result of the entire evolution of the star, determine the pattern
   of SN outburst in many respects.
The pre-SN model for SN 1999em has the structure of a red supergiant with
   a radius of 500 $R_{\sun}$, and the chemical composition of its outer layers
   is the standard solar composition.
It is quite definite that the progenitor of SN 1987A was the star Sanduleak
   $-69^{\circ}202$, a B3~Ia blue supergiant.
The pre-SN model for SN 1987A has a radius of 35 $R_{\sun}$, and
   the chemical composition of its outer layers is typical of the LMC chemical
   composition, $X=0.743$, $Y=0.251$, and $Z=0.006$ (Dufour \cite{dufour84}).
It is worth noting that a deficit of heavy elements in the LMC matter compared
   to the standard solar composition favors the formation of blue supergiants
   (Arnett \cite{arn87}; Hillebrandt et al. \cite{hhtw87}).
In the case of SN 1987A, a relative compactness of the pre-SN is a major
   factor in understanding the peculiar properties of this phenomenon (Grassberg
   et al. \cite{ginu87}).

\begin{table}[t]
\caption[]{Shock breakout and UV flash in SN 1999em and SN 1987A.}
\label{tab:sbsn99sn87}
\centering
\begin{tabular}{@{ } c @{ } c @{ } c @{ } c @{ } c @{ } c @{ } c @{ } c @{ } c @{ }}
\hline\hline
\noalign{\smallskip}
 SN & $t_{sh}$ & $T_c^{max}$ & $T_{eff}^{max}$ & $\log L_{bol}^{max}$ & $\Delta t_L$
       & $N^{tot}$ & $t^N$ & $E_{rad}^N$ \\
       & (days) & ($10^5$\,K) & ($10^5$\,K) & (erg\,s$^{-1}$) & (days)
       & ($10^{56}$) & (days) & ($10^{46}$\,erg) \\
\noalign{\smallskip}
\hline
\noalign{\smallskip}
99em & 0.865 & 3.84 & 1.76 & 44.81 & 0.0200 & 276.8 & 1.234 & 159 \\
87A  & 0.049 & 13.2 & 5.68 & 44.63 & 0.0007 & 7.615 & 0.266 & 6.24 \\
\noalign{\smallskip}
\hline
\end{tabular}
\end{table}
The basic characteristics of the shock breakout and the UV flash in SN 1999em
   and SN 1987A are summarized in Table~\ref{tab:sbsn99sn87} where
   $t_{sh}$ is the time when the shock wave reaches the stellar surface,
   $T_c^{max}$ the maximum of the color temperature,
   $T_{eff}^{max}$ the maximum of the effective temperature,
   $L_{bol}^{max}$ the maximum of the bolometric luminosity,
   $\Delta t_L$ the width of the luminosity peak at a half level of its
   maximum, $N^{tot}$ the total number of ionizing photons above 13.598 eV
   for the whole outburst, $t^N$ the time when the number of ionizing photons
   measures up $90\%$ of the total number, and $E_{rad}^N$ the energy
   radiated by the SN during this time.

Under the comparable ejecta masses and the explosion energies, the smaller
   the pre-SN radius, the higher the velocity of matter in the
   outermost layers with a sharp decline of density where the matter accelerates
   due to the effect of hydrodynamic cumulation.
At the same time the characteristic time of the envelope expansion becomes shorter
   and the adiabatic losses of energy greater.
Clearly, in SN 1999em the shock wave reaches the stellar surface much later than
   in SN 1987A, and the color and effective temperatures jump to lower values,
   while the bolometric luminosity rises to a higher value and its peak is
   much wider (Table~\ref{tab:sbsn99sn87}).
Note that in SN 1999em the bolometric luminosity peak coincides with the maximum
   of the effective temperature and occurs after the shock wave reaches the
   stellar surface in contrast to SN 1987A in which the maximum of the effective
   temperature precedes the peak, and the shock wave emerges between them on the
   pre-SN surface.
The UV flash in SN 1999em produces a greater total number of ionizing photons
   by a factor of 36.3, and it lasts longer than in SN 1987A by a factor of 4.6,
   the radiated energy being greater by a factor of 25.5
   (Table~\ref{tab:sbsn99sn87}).

\begin{figure}[t]
   \resizebox{\hsize}{!}{\includegraphics{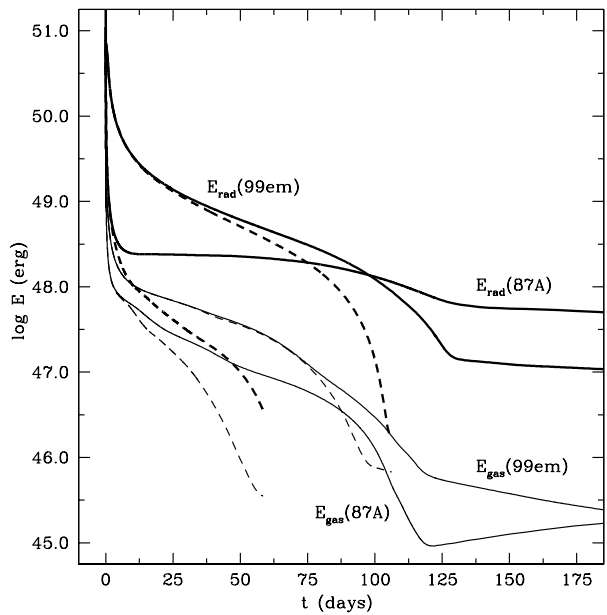}}
   \caption{%
   Total radiation (\emph{thick solid line\/}) and gas (\emph{thin solid line\/})
      energies as a function of time for SN 1999em and SN 1987A.
   The dependence on the energy deposition from radioactive decays is shown by the
      special hydrodynamic models without $^{56}$Ni in the envelope
      (\emph{dashed line\/}).
   }
   \label{fig:engsn99sn87}
\end{figure}
\begin{figure}[t]
   \resizebox{\hsize}{!}{\includegraphics{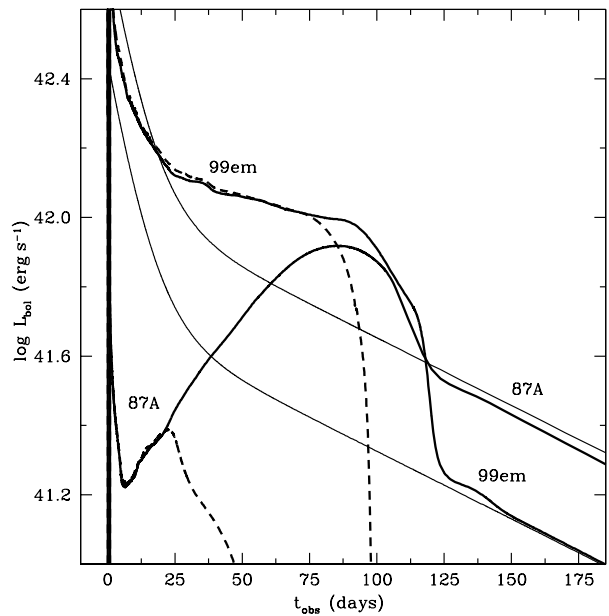}}
   \caption{%
   Comparison of the calculated bolometric light curve (\emph{thick solid
      line\/}) with the gamma-ray luminosity (\emph{thin solid line\/})
      for SN 1999em and SN 1987A.
   The special hydrodynamic models containing no $^{56}$Ni in the envelope
      show the influence of the energy deposition from radioactive decays
      (\emph{dashed line\/}).
   }
   \label{fig:lumsn99sn87}
\end{figure}
The smaller the pre-SN radius, the greater the adiabatic losses of energy
   and the shorter the CRW phase --- a characteristic feature of SNe IIP.
A difference between the explosions of the red and blue supergiants is radical
   in the total radiation and gas energies as a function of time
   (Fig.~\ref{fig:engsn99sn87}) and, especially, in the light curves
   (Fig.~\ref{fig:lumsn99sn87}).
To clarify the physical processes in SN 1999em and SN 1987A, we compare the optimal
   models of these SNe with the special hydrodynamic models containing no
   $^{56}$Ni in the envelope.
Note that during the entire outburst the total radiation energy is
   much greater than the total gas energy of the envelope.
A lack of $^{56}$Ni makes the energy losses due to the adiabatic expansion
   a dominant process controlling the energy balance of an SN.
In this case, the blue supergiant structure of the pre-SN for SN 1987A results
   in such huge energy losses that its total radiation energy is much less than that
   of SN 1999em the whole time and is completely exhausted in about 50 days, while
   for SN 1999em this happens later by around 50 days.
Accordingly, the luminosity at the CRW phase is maintained by the stored energy
   during only 22 days for SN 1987A and about 75 days for SN 1999em.

The subsequent run of the bolometric light curves in the optimal models of these
   SNe is entirely determined by the energy deposition of gamma rays from the
   radioactive decays and shows the key role played by $^{56}$Ni and $^{56}$Co
   in powering the light curve (Fig.~\ref{fig:lumsn99sn87}).
The luminosity of SN 1999em exceeds the gamma-ray luminosity from the peak to
   the radioactive tail.
On the contrary, the luminosity of SN 1987A drops abruptly well below the gamma-ray
   luminosity during the adiabatic cooling phase, grows to the maximum
   after the CRW phase in about 65 days exceeding the latter at day 60, and forms
   a wide dome on the light curve, declining to the radioactive tail and subsequently
   diving under it.
The wide dome on the light curve of the SN 1987A is physically equivalent to the
   phases of the radiative diffusion cooling, the exhaustion of radiation energy,
   and the plateau tail in the evolution of SN 1999em in the sense that it
   contains the basic features of all these phases.

At the stage of the radioactive tail, there is a striking difference between
   SN 1999em and SN 1987A: the internal gas energy of the envelope decreases
   with time in SN 1999em and increases in SN 1987A (Fig.~\ref{fig:engsn99sn87});
   the bolometric luminosity of SN 1999em coincides with the corresponding
   gamma-ray luminosity, while the bolometric luminosity of SN 1987A runs below
   it (Fig.~\ref{fig:lumsn99sn87}).
The gas energy is controlled by the energy rate of gamma-ray deposition,
   the energy losses due to the adiabatic expansion, and the net rate of
   absorption-emission processes.
At a given phase the last is dominated by the emission processes.
In the envelope of SN 1999em, the adiabatic energy losses are much greater than
   those in the envelope of SN 1987A where the thermal gas energy is much less
   than that of SN 1999em.
This fact explains both the decrease in the total gas energy with time in
   SN 1999em and its increase in SN 1987A.

The radiation energy is ruled mainly by the energy rate of gamma-ray deposition
   and the rate of work done by radiation pressure.
In the envelope of SN 1999em the total radiation energy is much less than that
   in the envelope of SN 1987A, and, consequently, the rate of work done by
   radiation pressure in SN 1999em is much less than that in SN 1987A as well.
It leads to a negligible difference between the bolometric and gamma-ray light
   curves in the optimal model of SN 1999em and to the bolometric light curve
   running below the gamma-ray light curve in the optimal model of SN 1987A.
Note that the higher a rate of work done by radiation pressure in an SN
   envelope, the larger the excess of the gamma-ray luminosity over the bolometric
   one at the radioactive tail when the SN envelope still remains
   optically thick for gamma rays.
In the optimal model of SN 1987A, the excess is nearly $7\%$
   (Fig.~\ref{fig:lumsn99sn87}), which corresponds to an underestimation of
   the $^{56}$Ni amount of the same value in equating the observed bolometric
   luminosity to the gamma-ray one.
Such an excess of the gamma-ray luminosity over the bolometric one is of great
   interest and should be taken into account in measuring the $^{56}$Ni amount
   from the observed bolometric luminosity at the radioactive tail.

It is very important that the approximation of homologous expansion may be used in
   the atmosphere models starting from nearly day 2.8 (Sect.~\ref{sec:dmd-free})
   for SN 1999em, a normal SN IIP, and from day 1 (Utrobin \cite{utr04}) for
   SN 1987A, a peculiar SN IIP.
Because of higher expansion velocities in the ejecta, the effects of expansion
   opacity and limb darkening are more prominent in SN 1987A than in SN 1999em.
For both SNe IIP, the crucial role of the time-dependent approach in atmosphere
   models at the photospheric epoch is evident from a comparison of spectral lines
   computed in this approach with those in the steady-state approximation.

\section{Discussion}
\label{sec:disc}
The aim of this paper is to explore the SN 1999em event, a normal SN IIP, by comparing
   the hydrodynamic models and the time-dependent atmosphere models with
  \emph{both} the photometric \emph{and} spectroscopic observations.
The resulting optimal model of SN 1999em succeeds in reproducing the observed
   features.
The bolometric light curve and the spectral evolution of the H$\alpha$ line are
   consistent with a radius of the pre-SN of $500 R_{\sun}$, a mass of
   the ejected envelope of $19 M_{\sun}$, an explosion energy of
   $1.3\times10^{51}$ erg, and a mass of radioactive $^{56}$Ni of
   $0.036 M_{\sun}$, the bulk of which is confined to layers ejected with
   velocities less than $\approx 660$ km\,s$^{-1}$.

The observational data of an SN are measured with some errors that yield
   the uncertainties in values inferred from them.
In the case of SN 1999em, we can translate these measurement errors of the
   observed parameters into uncertainties of the physical parameters using the
   approximate formulae (\ref{eq:logr}--\ref{eq:loge}).
To assign an uncertainty to the bolometric luminosity in the middle of the
   plateau, we adopt a relative error of $7\%$ for measuring the bolometric
   luminosity itself (Elmhamdi et al. \cite{edc03}) and an uncertainty of
   $\pm1$ Mpc for the Cepheid distance to the host galaxy (Leonard et al.
   \cite{lknt03}).
The resulting relative error in the bolometric luminosity in the middle of the
   plateau is $26\%$, as well as the error in the mass of radioactive $^{56}$Ni.
The relative errors in the photospheric velocities do not exceed $5\%$,
   inferred from the Fe II 4924, 5018, and 5169 \AA\ absorption lines in the
   middle of the plateau (Hamuy et al. \cite{hpm01}; Leonard et al. \cite{lfg02}).
The scatter in the explosion date yielded by the EPM varies within $\pm2$ days
   (Hamuy et al. \cite{hpm01}; Leonard et al. \cite{lfg02}; Elmhamdi et al.
   \cite{edc03}).
To estimate the measurement errors in the characteristic duration of the light
   curve, we consequently adopted an uncertainty of $\pm4$ days.
All the accepted errors in the observed values yield the following
   uncertainties in physical parameters: $\pm200 R_{\sun}$ in the pre-SN
   radius, $\pm1.2 M_{\sun}$ in the ejecta mass, $\pm0.1\times10^{51}$ erg in
   the explosion energy, and $\pm0.009 M_{\sun}$ in the mass of radioactive
   $^{56}$Ni.

Elmhamdi et al. (\cite{edc03}) note a clear flattening in the BVRI light curves
   of SN 1999em after the main plateau at the beginning of the radioactive tail
   and call it ``the second plateau''.
The ``UBVRI'' bolometric light curve of SN 1999em consequently reflects the second
   plateau.
The observed second plateau coincides in time with the plateau tail of the bolometric
   light curve for the optimal model (Sect.~\ref{sec:dmd-secnd}), but runs slightly
   below a line of the radioactive tail in contrast to the latter.
It seems that the real bolometric luminosity of SN 1999em is
   greater than the ``UBVRI'' luminosity during the second plateau.
Elmhamdi et al. (\cite{edc03}) also report on the second plateau on the tail
   for SN 1991G and SN 1997D, and find that its duration is correlated with
   the amount of ejected $^{56}$Ni, as well as the duration of the plateau tail
   does.
Of course, a second plateau nature that is different from that of the plateau tail
   cannot be excluded at the present time.

In the same study, Elmhamdi et al. explore the width and position of the He I
   10\,830 \AA\ line, which is sensitive to the nonthermal ionization and excitation
   produced by the radioactive decays, and show that in the ejecta of SN 1999em
   the bulk of $^{56}$Ni was distributed inside a region with a velocity less than
   1100 km\,s$^{-1}$ in the close hemisphere.
Obtained in the optimal model, a weak $^{56}$Ni mixing in velocity space up to
   $\approx 660$ km\,s$^{-1}$ agrees well with this $^{56}$Ni distribution.
By interpreting the temporal evolution of the [O I] 6300, 6364 \AA\ doublet profile
   in terms of the dust formation, they conclude that the dust occupies a sphere
   with a velocity of $\approx 800$ km\,s$^{-1}$.
Dust grains are formed by heavy elements, so we should estimate the volume
   containing them.
The characteristic velocity of the outer edge of the metal-rich region is
   nearly $850$ km\,s$^{-1}$ in the optimal model.
Moreover, the $^{56}$Ni bubble shell containing heavy elements is unstable to
   the Rayleigh-Taylor instability and, consequently, should be broken by it
   (Basko \cite{basko94}) and mixed, thereby producing dense clumps in which
   dust can form.
Thus, inside the sphere of a velocity of roughly $850$ km\,s$^{-1}$ in the
   optimal model of SN 1999em, there are conditions that favor the dust formation
   in the ejecta at late times.
In addition, from the observed [O I] 6300, 6364 \AA\ doublet luminosity, they
   infer the oxygen mass of $\sim 0.3-0.4 M_{\sun}$, which is consistent with
   the oxygen mass of $0.53 M_{\sun}$ estimated for the optimal model.

Baklanov et al. (\cite{bbp05}) carry out the hydrodynamic study of SN 1999em
   in multi-group approximation, compare the calculated hydrodynamic model to
   both the observed UBVRI light curves and the photospheric velocity, and advocate
   the hydrodynamic model for the distance of 7.5 Mpc with the pre-SN radius
   of 450 $R_{\sun}$, the ejecta mass of 15 $M_{\sun}$, and the explosion energy
   of $0.7\times10^{51}$ erg.
However, the low metallicity of $Z=0.004$ for a red giant pre-SN and the
   uniform chemical composition with $X=0.7$ and $Z=0.004$ from the surface
   to its center are questionable.
It is worth noting that their model for the distance of 12 Mpc is not far from
   our optimal model in the basic parameters: the pre-SN radius
   of 1000 $R_{\sun}$, the ejecta mass of 18 $M_{\sun}$, and the explosion energy
   of $10^{51}$ erg.

The ejected envelope of $19 M_{\sun}$ and a neutron star of $1.58 M_{\sun}$
   in the optimal model of SN 1999em correspond to a $20.58 M_{\sun}$
   pre-SN star.
Assuming an extremely low mass loss, we may now conclude that the lower mass
   limit of the progenitor of SN 1999em on the main sequence is nearly
   $21 M_{\sun}$.
Heger et al. (\cite{hlw00}) report the results of stellar evolution
   calculations of non-rotating main-sequence stars in the mass range of
   $10-25 M_{\sun}$.
It turns out that a main-sequence star of $25 M_{\sun}$ evolves to the
   pre-SN star of $18.72 M_{\sun}$ with the final helium core mass
   of $7.86 M_{\sun}$.
A linear extrapolation for the pre-SN mass of $20.58 M_{\sun}$ yields
   the stellar mass of $28.9 M_{\sun}$ on the main sequence and the final helium
   core mass of $9.6 M_{\sun}$.
Thus, the pre-SN of SN 1999em most likely developed from a main-sequence
   star in the mass range of $21-29 M_{\sun}$.

Generally speaking, there is an opportunity to put constraints on the masses and
   the evolutionary states of the progenitors of core collapse SNe from
   direct observations.
But it is a difficult task to identify the progenitors of SNe, and it is
   impossible to predict the time of explosion of a massive star precisely.
For these reasons Smartt (\cite{s02}) suggests estimating the initial masses of
   the progenitors by directly identifying SNe progenitors in the
   archive images of outburst sites taken prior to the explosions, by estimating
   the bolometric luminosity limits of progenitors as a function of stellar
   effective temperature, and by comparing them with stellar evolutionary tracks
   on the Hertzsprung-Russell diagram.
Applying this method to SN 1999em, Leonard et al. (\cite{lknt03}) have derived the upper
   mass limit for the progenitor of $20\pm5 M_{\sun}$ for the Cepheid distance
   to NGC 1637, and Smartt et al. (\cite{smg03}) obtained the upper limit of
   $15 M_{\sun}$ for the distance of 11 Mpc.
Evidently, our estimate of the progenitor mass is in good agreement with the
   first upper mass limit.

It should be emphasized that the estimates of the bolometric luminosity for
   progenitors are hardly affected by the circumstellar dust, and a comparison
   with stellar evolutionary tracks is far from being unambiguous as
   the well-known case of SN 1987A demonstrated.
There is growing observational evidence of the crucial role played by circumstellar
   dust in evaluating the bolometric luminosity of a progenitor star.
For example, Barlow et al. (\cite{bsf05}) show that a large fraction of
   the 5 mag of extinction within the host galaxy NGC 6946 toward SN 2002hh
   might be due to circumstellar gas dust, which might condense within a stellar
   wind from an earlier M supergiant or luminous blue variable phase of the evolution
   of the progenitor star.
And Massey et al. (\cite{mpl05}) find that a significant fraction of red
   supergiants in Galactic OB associations and clusters show up to several
   magnitudes of excess visual extinction compared to OB stars in the same
   regions and argue that this is very likely due to circumstellar dust around
   the red supergiants.
Note that the existence of circumstellar dust before an SN outburst does
   not contradict the lack of influence from dust on the photometric properties
   of an SN.
Dust grains are most likely evaporated by the intense radiation flash
   during the shock breakout inside a sphere of radius of $\sim 0.1$ pc
   (Dwek \cite{dwek83}).

A good fit of the calculated bolometric light curve for the optimal model to
   that observed for SN 1999em is achieved with mutual mixing of the
   hydrogen-rich and helium-rich matter at the helium/hydrogen composition
   interface.
The sharp boundary between the hydrogen-rich and helium-rich layers
   (Fig.~\ref{fig:crchcm}a) at the edge of the helium core --- a characteristic of
   the evolutionary models of pre-SNe --- changes the light curve producing
   a distinct bump at the end of the plateau (Fig.~\ref{fig:crchcm}b).
Note that the same feature was obtained by Chieffi et al. (\cite{cdhls03})
   in modelling the theoretical light curves of SNe IIP with the evolutionary
   models of pre-SNe.
In addition to a deep hydrogen mixing in velocity space downward to $\approx 700$
   km\,s$^{-1}$, the optimal model of SN 1999em is characterized by a weak
   $^{56}$Ni mixing up to $\approx 660$ km\,s$^{-1}$ (Fig.~\ref{fig:chcom} and
   Table~\ref{tab:hmsn99sn87}).

The hydrodynamic model for SN 1987A is, in turn, characterized by a moderate
   $^{56}$Ni mixing in velocity space up to $\sim 3000$ km\,s$^{-1}$ and a deep
   hydrogen mixing downward to $\sim 600$ km\,s$^{-1}$ (Utrobin \cite{utr05}).
Moreover, the observations of SN 1987A provide clear evidence of moderate
   mixing of the bulk of radioactive $^{56}$Ni up to velocity $\sim$~3000 km\,s$^{-1}$
   and deep hydrogen mixing down to $\sim$~500 km\,s$^{-1}$.
For example, the [Ni~II] 6.64 $\mu$m profile at day 640 gives a velocity
   $v_{\mathrm{FWHM}}=3100$ km\,s$^{-1}$ and its modelling results in
   a maximum velocity of 2600 km\,s$^{-1}$ (Colgan et al. \cite{chelh94}).
The fact that the H$\alpha$ profile on day 498 (Phillips et al. \cite{phhsk90})
   is not flat-topped implies that there is no large cavity free of hydrogen
   at the center of the envelope and that hydrogen is mixed downward to
   $\sim$~500 km\,s$^{-1}$.

Within the framework of the neutrino-driven explosion mechanism for core
   collapse SNe, Kifonidis et al. (\cite{kpsjm03}, \cite{kpsjm06})
   carried out two-dimensional simulations of explosion models with high-order
   mode perturbations and those with low-order ($l=2$ and $l=1$) unstable modes
   induced by the initial global deformation of the shock.
The low-mode explosion models exhibit final iron-group velocities of $\sim$~3300
   km\,s$^{-1}$ and deep hydrogen mixing downward to a velocity of $\sim$~500
   km\,s$^{-1}$, which are consistent with the observations of SN 1987A.
The moderate $^{56}$Ni mixing up to a velocity of $\sim$ 3300 km\,s$^{-1}$ is
   the result of the larger initial maximum velocities of metal-rich clumps,
   compared to the high-mode models, which protect the fastest clumps from
   the strong interaction with the reverse shock that forms below the
   helium/hydrogen composition interface.
The initial global deformation of the shock leads to the growth of the
   Richtmyer-Meshkov instability that in turn results in the strong inward
   mixing of hydrogen at the helium/hydrogen composition interface.

It is likely that the strongly anisotropic explosions are able to explain both
   weak and moderate $^{56}$Ni mixing, depending on the pre-SN structure and
   the character of the interaction between metal-rich clumps and the reverse shock.
Thus, the weak $^{56}$Ni mixing and the deep hydrogen mixing, which are the
   characteristic features of the optimal model for SN 1999em, cannot be excluded
   within the framework of the neutrino-driven explosion mechanism
   (Kifonidis \cite{k06}).
These promising capabilities make our interpretation of the SN 1999em event more
   physically grounded and confirm the suggestion of Chieffi et al. (\cite{cdhls03})
   that in reality the inner layers of the expelled envelope are strongly mixed
   during the explosion.

Based on the hydrodynamic models of SNe IIP, Litvinova \& Nadyozhin (\cite{ln83},
   \cite{ln85}) constructed simple approximate formulae for evaluating the initial
   radius, the ejecta mass, and the explosion energy --- three basic physical
   parameters --- from the observed properties of individual SN IIP.
Using these approximate formulae for SN 1999em and assuming the distance to
   the host galaxy to be 11.08 Mpc, Nadyozhin (\cite{n03}) derived the initial
   radius of 414 $R_{\sun}$, the ejecta mass of 15.0 $M_{\sun}$, and the explosion
   energy of $0.68\times10^{51}$ erg.
The difference in the ejecta mass and the explosion energy, compared
   to those of the optimal model, results mainly from neglecting the influence
   of radioactive $^{56}$Ni on the light curve and using the approximation of
   equilibrium radiative diffusion in their hydrodynamic models.

We go one step further to construct the approximate formulae
   (\ref{eq:logr}--\ref{eq:loge}), which include the influence of radioactive $^{56}$Ni
   on the light curve and are based on the hydrodynamic models computed in terms of
   radiation hydrodynamics in the one-group approximation with non-LTE effects in
   the average opacities and the thermal emissivity, with nonthermal ionization, and
   with the contribution of lines to opacity.
The resulting approximate formulae, obtained in space of the physical parameters, are
   valid only for the vicinity of the optimal model because of the sensitivity of the
   bolometric light curve to the pre-SN structure and to its chemical
   composition.
To construct the global, approximate relations covering a wider range of the
   parameters, we have to use a set of hydrodynamic models computed for
   \emph{many real} SNe IIP.
The optimal model for SN 1999em is the first element in this set.

\section{Conclusions}
\label{sec:concl}
We have presented a comprehensive study of a normal Type IIP SN 1999em.
Our main aim in this paper was to show the absolute necessity of the simultaneous
   interpretation of \emph{both} the photometric \emph{and} spectroscopic
   observations.
To do this, we calculated the hydrodynamic models of SN 1999em complemented
   by the atmosphere models with the time-dependent kinetics and energy balance.
We focused mainly on the optimal hydrodynamic model of SN 1999em and
   subsequent comparison to a peculiar Type IIP SN 1987A.
Our results can be summarized in the following conclusions.

\begin{itemize}

\item[--]
The bolometric light curve of SN 1999em and its spectral evolution of the H$\alpha$
   line are consistent with a radius of the pre-SN of $500\pm200 R_{\sun}$,
   a mass of the ejected envelope of $19.0\pm1.2 M_{\sun}$, an explosion energy of
   $(1.3\pm0.1)\times10^{51}$ erg, and a mass of radioactive $^{56}$Ni of
   $0.036\pm0.009 M_{\sun}$, the bulk of which is confined to layers ejected with
   velocities less than $\approx 660$ km\,s$^{-1}$.
The optimal hydrodynamic model with the helium core mass in the range of 5.6--8.1
   $M_{\sun}$ matches the observed bolometric light curve.

\item[--]
The adequate hydrodynamic and atmosphere models of SN 1999em distinguish between
   the short distance of 7.85 Mpc, the average value of the EPM distance estimates
   to the host galaxy, and the Cepheid distance of 11.7 Mpc.
They are inconsistent with the short distance of 7.85 Mpc, which should
   be discarded.

\item[--]
It is shown that the hydrogen recombination in the atmosphere of a normal Type
   IIP SN 1999em, as well as most likely other SNe~IIP, at the photospheric epoch
   is essentially a time-dependent phenomenon in accordance with the analysis
   of Utrobin \& Chugai (\cite{uc02}, \cite{uc05}).

\item[--]
This study reveals the time development of a normal SN~IIP with the
   following stages: a shock breakout, an adiabatic cooling phase, a phase of
   cooling and recombination wave, a phase of radiative diffusion cooling,
   an exhaustion of radiation energy, a plateau tail, and a radioactive tail.

\item[--]
The plateau tail of the theoretical bolometric light curve coinciding in time with
   the second plateau observed in SN 1999em (Elmhamdi et al. \cite{edc03}) runs
   slightly above a line of the radioactive tail in contrast to the latter.
It seems that the real bolometric luminosity of SN 1999em is
   greater than the ``UBVRI'' luminosity during the second plateau.
The duration of the plateau tail, as well as the second plateau, is correlated with
   the amount of ejected $^{56}$Ni.

\item[--]
The bolometric light curve depends weakly on the chemical composition
   in the outer layers of the ejecta beyond the helium core, while hydrogen
   is abundant there and controls the opacity of matter.
A mutual mixing of hydrogen-rich and helium-rich matter in the inner layers of
   the ejecta guarantees a good fit of the calculated bolometric light curve
   to what is observed.
Note that the sharp boundary between the hydrogen-rich and helium-rich layers
   at the edge of the helium core --- a characteristic of the evolutionary models
   of pre-SNe --- changes the light curve producing an unobserved bump
   at the end of the plateau.

\item[--]
A contribution of numerous metal lines to opacity plays a fundamental role
   in reproducing the observed light curve of SN 1999em.
The decrease in opacity due to disregarding the line contribution speeds up
   the radiation diffusion and causes the bolometric luminosity to increase
   significantly compared to what is observed during nearly the whole outburst
   and, as a consequence, causes the characteristic duration of the light curve
   to shorten by about 10 days.

\item[--]
The comparison of a normal Type IIP SN 1999em with a peculiar Type IIP SN 1987A
    reveals two very important results for SN theory.
First, the masses of helium cores in the pre-SNe of SN 1999em and SN 1987A
   are close enough to suppose that in the final stages of stellar evolution
   nearly the same iron cores form within the pre-SNe.
This fact and roughly the same explosion energies of SN 1999em and SN 1987A
   together imply a unique explosion mechanism for these core collapse SNe.
Second, the optimal model for SN 1999em is characterized by a weaker $^{56}$Ni
   mixing in velocity space up to $\approx 660$ km\,s$^{-1}$ compared to
   a moderate $^{56}$Ni mixing up to $\sim 3000$ km\,s$^{-1}$ in the case of
   SN 1987A, hydrogen being mixed deeply downward to $\approx 700$ km\,s$^{-1}$
   and $\sim 600$ km\,s$^{-1}$.

\item[--]
A significant excess of the gamma-ray luminosity over the bolometric one, shown
   by the optimal model of SN 1987A at the radioactive tail, is of great interest
   and should be taken into account in measuring the $^{56}$Ni amount from the
   observed bolometric luminosity at this stage when the SN envelope still
   remains optically thick for gamma-rays.
The higher the rate of work done by radiation pressure in an SN envelope,
   the larger an excess of the gamma-ray luminosity over the bolometric one at
   the radioactive tail.

\item[--]
It is shown that the approximation of homologous expansion may be used in the EPM
   and SEAM for determining the distances to normal SNe IIP starting from
   nearly day 3 after the SN explosion.

\item[--]
Based on the hydrodynamic models in the vicinity of the optimal model of SN 1999em,
   we derive the approximate relationships between the basic physical and
   observed parameters.
The basic physical parameters are the pre-SN radius, the ejecta mass,
   the explosion energy, and the total $^{56}$Ni mass.
The observed properties of SNe IIP are described by the three parameters: the
   characteristic duration of the light curve, the bolometric luminosity and
   the photospheric velocity in the middle of the plateau.
To cover a wider range of the parameters, beyond the vicinity of the optimal model,
   we have to construct a global approximation that should be based on a set of
   hydrodynamic models of \emph{many real} SNe IIP.

\end{itemize}

Finally, we emphasize that the simultaneous analysis of the photometric and
   spectroscopic observations is in fact crucial for correctly interpreting
   core collapse SNe and express the hope that the obtained basic
   parameters and properties of SN 1999em determine its evolutionary state
   and will be a reliable reference for the proper mechanism of SN
   explosions.

\begin{acknowledgements}
The author is grateful to Wolfgang Hillebrandt and Ewald M\"{u}ller for hospitality
   during his stay at the MPA.
The author would also like to thank Nikolai Chugai, Konstantinos Kifonidis, and
   Dmitrij Nadyozhin for many discussions, and the referee David Branch for
   helpful comments.
This work was supported in part by the Russian Foundation for
   Fundamental Research (04-01-17255).
\end{acknowledgements}


\end{document}